\documentclass[aps,prl,10pt,superscriptaddress,twocolumn,preprintnumbers,amsmath,amssymb]{revtex4-2}

\usepackage{mathtools}
\usepackage{slashed}
\usepackage{tensor}
\usepackage{bbold}  
\usepackage{amscd}
\usepackage{color}
\usepackage{slashed}

\usepackage{calrsfs}  
\DeclareMathAlphabet{\pazocal}{OMS}{zplm}{m}{n}

\allowdisplaybreaks

\newcommand{\tr}{\mathrm{tr}\,}

\newcommand{\dd}{\mathrm{d}}
 
\newcommand{\cgamma}{\underline{\gamma}\vphantom{\gamma}}

\usepackage[dvipsnames]{xcolor}

\definecolor{brickred}{rgb}{0.8, 0.25, 0.33}
\newcommand\myshade{85}
\colorlet{mylinkcolor}{BrickRed}
\colorlet{mycitecolor}{NavyBlue}
\colorlet{myurlcolor}{Aquamarine}

\usepackage{hyperref}
\hypersetup{
  linkcolor  = mylinkcolor!\myshade!black,
  citecolor  = mycitecolor!\myshade!black,
  urlcolor   = myurlcolor!\myshade!black,
  colorlinks = true,
}

\begin{document}
  
\title{Graphene Shapes from Quantum Elasticity}

\author{Pablo A. Morales}

\email{pablo$_$morales@araya.org}
\affiliation{Research Division, Araya Inc., Tokyo 107-6019, Japan}
\affiliation{Centre for Complexity Science, Imperial College London, London SW7 2AZ, UK}

\author{Pavel Castro-Villarreal}

\email{pcastrov@unach.mx}
\thanks{author to whom correspondence should be addressed.}
\affiliation{Facultad de Ciencias en Física y Matemáticas, Universidad Autónoma de Chiapas, Carretera Emiliano Zapata, Km. 8, Rancho San Francisco, C. P. 29050, Tuxtla Gutiérrez, Chiapas, México}

\begin{abstract}
Temperature constraints are highly desirable in the experimental setup when seeking the synthesis of new carbon structures. Fluctuations of the Dirac field result in temperature-dependent corrections to the Helfrich-Canham formulation, which governs the classical elasticity of the graphene membrane at equilibrium. Here, we examine the emergent shapes allowed by the effective model up to quadratic order in Ricci curvature and discuss the constraints required to observe them. We determine the mechanical stability conditions and provide a phase diagram characterized by the appearance of a critical temperature $T_{\rm c}$ that distinguishes between carbon nanotube and fullerene phases. The observation of minimal and developable surfaces is anticipated in the high- and low-temperature regimes, respectively. Additionally, a Beltrami trumpet surface is forecasted when the membrane is subjected to an external source balancing out internal Helfrich stresses.
\end{abstract}

\maketitle

\section{I. Introduction}
The rapid improvement in graphene synthesis techniques has increased interest in the problem of how morphology relates to its properties~\cite{RevModPhys.81.109,nano11071694}. Graphene can be bent to form corrugated graphene, folded to form fullerenes, rolled into carbon nanotubes, and stacked in the shape of graphite, making it the mother of graphitic materials~\cite{Novoselov2}. The production of flat graphene sheets is also subject to deformations; dislocations introduce strain fields that lead to instabilities~\cite{Naumis_2017,*Naumis_2024}, and these stresses are then relieved by out-of-plane ripples which manifest in the presence of experimentally observed nanobubbles~\cite{doi:10.1126/science.1191700}. Furthermore, the Dirac-like spectrum of low-energy excitations suggests the possibility of tabletop experiments as a test bed for relativistic quantum phenomena and vice versa. In this way, the coupling of massless Dirac fermions to the sample geometry enables us to directly probe morphological effects. Indeed, the emergence of pseudogauge fields from the ripples in graphene modulated by curvature~\cite{Castro2017,Morales2023} and observed in the range of 200~\cite{Kun2019} to 300\,T~\cite{doi:10.1126/science.1191700} critically affects the sample's electrical properties.

However, without an accurate tight-binding (TB) description of curved graphene, it becomes difficult to determine the precise field theory for low energies that allows us to model its properties. TB models for deformed graphene are usually based on position-dependent hopping integrals and a slightly deformed honeycomb lattice~\cite{Naumis_2017}. Curved graphene, however, generally ceases to be a crystal due to the changes in its structural lattice produced by the bending or stretching of graphene. Hence, it is not obvious how to formulate a TB model that accounts for the topological defects that make up curvature.  
Consistently taking into account the lattice gauge symmetry of standard TB models leads to nonlinear modifications to the Dirac dynamics in curved space-time~\cite{MattWiseman1,*MattWiseman2} (see~\cite{CommentPaper,ReplyToComment} for other views). Resolving the break of translational symmetry in crystalline structures precedes a proper, effective field theoretic formulation. Alternatively, a perhaps more phenomenological approach, considering the curved-space Dirac field theory is currently  the simplest way to investigate the electronic degrees of freedom in corrugated graphene. In fact, up to a fitting via density functional theory, the low-energy spectrum obtained with a TB model of a graphene sheet deformed by a Gaussian bump under a transverse magnetic field agrees well with the spectrum predicted from the curved Dirac model subject to the same conditions~\cite{Castro2024}.  

Grounding an experimental setup demands knowledge of the surface embedding and the response from the electronic degrees of freedom confined to the \textit{membrane} at a given temperature. Actually, the self-assembly formation of carbon surfaces like carbon nanotubes, fullerenes, and carbon nanocones, among other curved graphene surfaces, is a complex phenomenon. The experimental methods known to produce these structures, like arc discharge, thermal pyrolysis, and chemical vapor deposition, among others, have revealed an out-of-equilibrium process for their formation~\cite{SRISUMA20211373, Chuvilin2010}, which is beyond of the known theoretical approach. However, to achieve a piece of a curved graphene surface at an equilibrium temperature, the surface must be mechanically stabilized, considering the Dirac fermionic degrees of freedom. In fact, it has been shown that the relativistic Dirac degrees of freedom result in the tendency of the membrane to crumple~\cite{PhysRevLett.120.261601}, which seems to be consistent with the experimental observation of the transformation from graphene to fullerene~\cite{Chuvilin2010}. The classical membrane free energy is thus corrected by a significant quantum contribution that critically affects its stability. A connection to curved graphene would also require inclusion of a non-Abelian gauge field to account for topological defects that may emerge~\cite{kolesnikov2006continuum}; however, as a naive model, one may ask what the experimental conditions at which a given structure may be expected are. Basic notions of geometry and topology have led to the prediction of positively curved carbon nanostructures with unique properties that were later synthesized~\cite{Kroto,*Lijima}. Negative-curvature carbon materials, on the other hand, despite being proposed more than a decade prior to the synthesis of graphene~\cite{Terrones, TERRONES2010351}, have yet to be observed in the laboratory. Despite great expectations, the unknown mechanisms for its production and experimental synthesis have made its observation uncertain. 

In this paper, we investigate the role played by Dirac field thermal fluctuations in the effective spatial geometry in graphene. We provide a \textit{shape equation} for extremal configurations along with the general equilibrium conditions under which they can be achieved. Based on a naive model for graphene membrane, we discuss the characteristic scale of some of the carbon structures observed in the laboratory as well as the temperatures at which they should be observed.

\section{II. Graphene membranes}
Commonly the space-time geometry used in two-dimensional (2D) materials is modeled as a $2+1$ ultrastatic space-time with pseudo-Riemannian metric $\dd s^2 = -(\dd x^{0})^2 + g_{ab}\dd x^a \dd x^b$, where $a, b=1, 2$ are the local indices of the surface and the metric $g_{ab}$ is associated with the geometry of a two-dimensional surface $\Sigma$. The material sheet is embedded in three-dimensional (3D) Euclidean space and thus an extrinsic description for the sheet geometry is required. The embedding functions are introduced through the mapping ${\bf X}:\mathcal{D}\subset \mathbb{R}^{2}\to \Sigma\subset\mathbb{R}^{3}$, where $\mathcal{D}$ is some open set. Additionally, for the past two decades, it has been suggested that the Helfrich-Canham (HC) free energy~\cite{CANHAM197061, *Helfrich1973} could serve as a suitable model for the elasticity of graphene sheets \cite{Kim_2008, RevModPhys.81.109}. 
The HC model is defined by the energy functional $H\left[{\bf X}\right]=\int \dd^{2}x\sqrt{g}\left[\frac{\alpha}{2}K^2+\kappa_{G}R+\sigma\right] 
$~\cite{DESERNO201511}, where $K=g^{ab}K_{ab}$ is twice the mean curvature, with $g^{ab}$ being the inverse metric tensor, $g$ being the metric determinant, and $K_{ab}$ being the extrinsic curvature tensor. $R$ is the Ricci curvature (twice the Gaussian curvature) related to the extrinsic curvature by the Gauss-Codazzi equation, $R=K^2-K_{ab}K^{ab}$, which, by Gauss's \textit{Theorema Egregium}, depends solely on the metric tensor $g_{ab}$ \cite{DoCarmo}. The energy functional consists of three terms: the $\alpha>0$ term is the bending energy; the $\kappa_{G}$ term is the Gaussian bending energy; and the $\sigma>0$ term is a surface energy or tension in the soft-matter literature, a Lagrange multiplier to fix the overall area of the membrane~\cite{DESERNO201511}. In this sense, $\alpha$ and $\kappa_{G}$ are the only phenomenological parameters of the HC model. Furthermore, for a compact surface $\Sigma$, the second term $\int \dd^{2}x\sqrt{g}R$ is a topological invariant known as the Euler characteristic $\chi(\Sigma)$.
The HC model offers a geometric perspective for graphene elasticity through the bending energy~\cite{YujieWei}. Recently, the first experimental evidence of the HC model was observed via atomic force microscopy~\cite{Ashino1, *Ashino2} by direct topological manipulation of the sheet and measurement of the elastic coefficients.

The electronic degrees of freedom on graphene in our approach, are modeled by chiral fermions in curved space-time $S[\bar{\Psi}, \Psi] = iv_{F}\int \dd ^{3}x\sqrt{g}\bar{\Psi}\slashed{D}\Psi$~\cite{IORIO2012, *Iorio2014}, where $v_{F}$ is the Fermi velocity associated with 2D flat Dirac materials. In this action, $\bar{\Psi}=\Psi^{\dagger}\gamma^{0}$, and $\slashed{D}=\cgamma^{\mu}\nabla_{\mu}$ is the usual Dirac operator, with $\cgamma^{\mu}=e^{\mu}_{\ell}\gamma^{\ell}$ being curved $\gamma$-matrices and $e^{\mu}_{\ell}(x)$ being vielbein fields. $\nabla_{\mu}=\partial_{\mu}+\Omega_{\mu}$ is the spinorial covariant derivative and $\Omega_{\mu}$ is its spin connection; for further details on the Hamiltonian description of graphene see Appendix~A. Dirac fields confined to the surface may, in principle, provoke modifications to the geometry of the membrane through thermal fluctuations of the fermion gas in the material.
To estimate the extent of this effect in the geometry, we pose a correction to the elastic bending energy, $H_{\rm eff}\left[{\bf X}\right]= H\left[{\bf X}\right]+\delta H_{\rm fermion}\left[{\bf X}\right]$, where, assuming the valley symmetry, $\delta H_{\rm fermion}\left[{\bf X}\right]=-\frac{g_{v}g_{s}}{\beta} \log Z\left(\beta, g\right)$ is the effective action from the Dirac fields with $g_{v}$ and $g_{s}$ being the valley and spin degeneracies and $\beta$ being the inverse of the thermal energy $k_{B}T$, where $k_{B}$ is the Boltzmann constant and $T$ is the temperature. To account for the thermal fluctuation of the Dirac field, we adopt a finite-temperature approach~\cite{Altland} by performing a Wick rotation of the Dirac operator in 2+1 dimensions with $\tau=-ix^{0}/(\hbar v_{F})$, i.e., $-i\hbar v_{F}\gamma^{0}\slashed{D}=\partial_{\tau}+\pazocal{H}$, with $\pazocal{H}=-i\hbar v_{F}\gamma^{0}\underline{\gamma}^{a}\nabla_{a}$. Furthermore, the Dirac field partition function $Z\left(\beta, g\right)$ is
\begin{equation}
    Z\left(\beta, g\right)=\int \mathcal{D}\Psi^{\dagger}\mathcal{D}\Psi e^{-\int_{0}^{\beta} \dd\tau \int_{\Sigma} \dd^{2}x \sqrt{g} \Psi^{\dagger}\left(\partial_{\tau}+\pazocal{H}\right)\Psi},\label{partitionfunction}
\end{equation}
with $\Psi^{\dagger}$ and $\Psi$ defined on the product manifold $S^{1}\times \Sigma$. It is worth mentioning that one can interpret $\pazocal{H}$ as the quantum Hamiltonian for a Dirac particle moving in the presence of a magnetic vector field potential $\Omega_{a}$ with a tensorial and space-dependent effective Fermi
velocity $v^{{\rm eff}, a}_{\ell}(x)=v_{F}e^{a}_{\ell}(x)$. Indeed, this is how a pseudomagnetic field emerges for the Dirac particles in a curved manifold \cite{Castro2017}.

\section{III. Effective free energy of a 2D Dirac material}

The fermionic path integral~\eqref{partitionfunction} at finite temperature {(see Appendix~B)} is,
\begin{eqnarray} \label{eq:finTfermZ}
    Z=\prod_{n\in\mathbb{Z}}{ \det}_{\mathbb{H}}\left\{\left(i\omega_{n}\gamma^{0}+\slashed{D}_{E}\right)\right\},
\end{eqnarray}
where $\omega_{n}$ are the fermionic Matsubara frequencies, and $\slashed{D}_{E}=\gamma^{0}\hat{\pazocal{H}}=i\underline{\gamma}^{a}\nabla_{a}$ is a Euclidean Dirac operator, and $\mathbb{H}$ is the Hilbert space associated with spinorial functions on the curved surface $\Sigma$. We adopt the signature $(-1,1,1)$ so that $\left(\gamma^{0}\right)^{\dagger}=-\gamma^{0}$ and $\left(\gamma^{a}\right)^{\dagger}=\gamma^{a}$; thus, the operator $i\omega_{n}\gamma^{0}+\slashed{D}_{E}$ is self-adjoint. From the identity $\log \det_{\mathbb{H}} \mathcal{O}={\rm Tr}_{\mathbb{H}}\log\mathcal{O}$ and recalling that $\log \lambda=-\int_{0}^{\infty}\frac{\dd s}{s}e^{-\lambda s}$ up to a divergent constant, \eqref{eq:finTfermZ} may be rewritten as, 
\begin{eqnarray}
\label{eq:logZ}
    \log Z=-\frac{1}{2}\int_{0}^{\infty}\frac{\dd s}{s}\left(\sum_{n\in\mathbb{Z}}e^{-\omega^{2}_{n}s}\right)\int_{ \Sigma}\dd A~K(s;\, x, x), 
\end{eqnarray}
where $dA=\dd^{2}x \sqrt{g}$ is the surface area element and $K(s; x, x^{\prime})=\left<x\left|e^{- Ds}\right|x^{\prime}\right>$ is the heat kernel of the operator $D=(\hbar v_F)^2\slashed{D}_{E}^2$ on $\Sigma$~\cite{parker2009quantum, VASSILEVICH2003279}. To compute $\log Z$ let us focus on its integrand. The heat kernel may be expanded in geometric invariants $E_k$ as 
\begin{equation}
    K(s;x,x) = \frac{1}{4\pi}\sum_{k\geq 0} (s\hbar^2 v_{F}^2)^{k-1} \tr\!(E_{k}), \label{eq:kernel_exp}
\end{equation}
with $\tr$ being the pseudospin trace. The first coefficients $E_k$ on a manifold without boundaries are local $O(2)$ invariant quantities~\cite{parker2009quantum}, (see Appendix~C). In two dimensions, the Ricci scalar curvature $R$ is the only independent component of the Riemann tensor 
$R_{abcd}=\frac{R}{2}(g_{ac}g_{bd}-g_{ad}g_{bc})$,  
allowing us to evaluate $E_k$ in terms of $R$, 
\begin{equation}
    E_{0}=\mathbb{1}, E_{1}=-\frac{\mathbb{1}}{12}R, E_{2}=-\frac{\mathbb{1}}{120}\left( \Delta_{g}R+\frac{1}{2}R^2\right).
\end{equation}

The Matsubara sum in~\eqref{eq:logZ} may be recast in terms of the Jacobi theta function $\vartheta_4$, and placing~\eqref{eq:kernel_exp} into $\log Z$,

\begin{equation}
    \label{eq:temp2}
    -\frac{1}{2}\int_{0}^{\infty}\frac{\dd s}{s}\vartheta_{4}\left[\tfrac{i\ell_{T}^2}{4\pi s}\right] \sum_{k\geq 0} \frac{s^{k-\tfrac{3}{2}} }{(4\pi)^{3/2}\ell_{T}^{-1}}\int_{\Sigma}\dd A~ \tr\!(E_{k}),
\end{equation}
{with} $\ell_{T}=\hbar v_{F}/k_{B}T$ corresponding to the effective thermal wavelength. Its zero temperature limit may be recovered from~\eqref{eq:temp2} while noting that $\vartheta_{4}\to 1$, consistent with the findings in~\cite{PhysRevD.46.5458}. Hence, from~\eqref{eq:temp2}, defining the heat kernel expansion coefficients at finite temperature as $a_{k}(x;\ell_{T}^2/s)$,
\begin{equation}
    -\frac{1}{2}\int_{0}^{\infty}\frac{\dd s}{s}\int_{S^1 \times \Sigma}\frac{\dd V}{(4\pi)^{3/2}}\sum_{k\geq 0} a_{k}(x;\ell_{T}^2/s) s^{k-\tfrac{3}{2}}, \nonumber
\end{equation}
the coefficients are now related to those at zero temperature as $a_{k}(x;\ell_{T}^2/s)=(\hbar v_F) \vartheta_{4}[i\ell_{T}^2 /4\pi s] \tr\!(E_k)$, with the Jacobi function carrying the temperature dependence. The divergent part of the effective action is contained in $s \to 0$. At this limit, however, the expansion coefficients are exponentially suppressed, converging to the zero temperature value.
Thus, finite temperature does not modify the divergent structure so the same counterterms suffice for renormalization. To capture the IR divergence in~\eqref{eq:temp2} we introduce a UV cutoff $\Lambda^{-2}$ and integrate up to some constant $s_0$ which may be taken to be arbitrary small. As argued above, at $s \to 0$ the coefficients are those at zero temperature, allowing for the explicit integration of $s$. Disposing of the $s_0$ term, the divergent part reads
\begin{equation}
    -\frac{1}{\ell_T} \int_{\Sigma}\frac{\dd A}{(4\pi)^{3/2}}\sum_{k = 0}^{[3/2]}\tr\!(E_k) \frac{\Lambda^{3-2k}}{2k-3}, \label{eq:counterterms}
\end{equation}
leading to a counterterm effective Lagrangian density $c_1 \Lambda^3 + c_2 R \Lambda$ with the constants $c_{1,2}$ determined from~\eqref{eq:counterterms}.
Rescaling $s \to \ell_{T}^{-2}s$, the renormalized free energy becomes
 
\begin{equation}
    F_{\mathrm{ren}}\left[{\bf X}\right]= \frac{1}{2\beta} \sum_{k \ge 0} g^{\mathrm{ren}}_k \ell_{T}^{2k-2} \int_{\Sigma} \frac{\dd A}{4\pi} \tr\! (E_{k}),
\end{equation}
{with} $g^{\mathrm{ren}}_k$ constants determined by the Mellin transform of the Jacobi theta function,
\begin{equation}\label{eq:ren_coeff}
    g^{\mathrm{ren}}_k \coloneqq \frac{1}{\sqrt{4\pi}}\int_{0}^{\infty} \dd s \,s^{k-\tfrac{5}{2}}\left(\vartheta_{4}\left[\tfrac{i}{4\pi s}\right]-\nu_k\right),
\end{equation}
where $\nu_k$, defined as $\nu_{0,1}=1$ and $\nu_{k \ge 2}=0$, encapsulates the effect of the renormalization procedure. The fermionic contribution to the free energy is $\delta H_{\rm fermion}\left[{\bf X}\right] = g_v g_s F_{\mathrm{ren}}\left[{\bf X}\right]$ with coefficients obtained analytically; $g^{\rm ren}_{0} = -3\zeta(3)$, $g^{\rm ren}_{1} = -2 \log (2)$, and $g^{\rm ren}_{2} = 1/4$. For $k \ge 2$, in general, $g^{\rm ren}_{k}= 2 \pi^{2-2k}[1-4^{1-k}]\Gamma (k-1)\zeta(2k-2)$ are monotonically decreasing. A strong curvature regime would require us to include higher order contributions, but they are quickly suppressed by the increasing powers of $\ell_{T}$ limited by the validity of the Dirac model; for most relevant systems $k>2$ contributions can be neglected.   

The effective Hamiltonian obtained after integrating out the Dirac degrees of freedom and performing the renormalization procedure has the following structure:
\begin{equation}\label{eq:eff_Ham}
\!H_{\rm eff}\left[{\bf X}\right]=\int_{\Sigma} \dd A\left[\frac{\alpha}{2}K^2+\kappa_{G}^{ \rm eff}R+\sigma_{\rm eff}+\frac{1}{2}\kappa^{{}_{(2)}}_{G}R^2\right],
\end{equation}
where $\sigma_{\rm eff} \coloneqq \sigma + \delta \sigma_{\rm eff}$ and $\kappa_{G}^{\rm eff} \coloneqq \kappa_{G} + \delta \kappa_{G}^{\rm eff}$; i.e., with the exception of the bending rigidity coefficient $\alpha$, all coefficients are modified by one-loop quantum corrections of the fermionic sector. The surface tension and Gaussian elastic module receive a temperature-dependent contribution  from the IR dynamics: $\beta \delta \sigma_{\rm eff} =-3 g_v g_s \zeta(3)/\ell_{T}^{2}$ and $\beta \delta \kappa_{G}^{\rm eff} = \frac{1}{6}g_v g_s\log(2)$, respectively. The sign of $\delta \sigma_{\rm eff}$ associated with vacuum energy is worth noting; this contribution will manifest in temperature constraints for the observation of carbon structures, which will be discussed later. In addition, $\delta \sigma_{\rm eff}$ is in agreement with the known expression for the heat capacity of graphene in the Dirac approximation~\cite{Castro2017}, as well as the Casimir-type contribution from the finite temperature calculation~\cite{fischetti2020does}. In contrast, an $R^2$ term has been induced with the emergent quantum elastic coefficient $\beta \kappa_{G}^{{}_{(2)}}= -\frac{g_v g_s}{960 \pi}\ell_{\mathrm{T}}^2$. This term is a consequence of the general expression obtained by the heat kernel expansion in two dimensions~\cite{VASSILEVICH2003279} considered in~\cite{fischetti2020does} to determine whether the sphere is a {\it global maximum} of their free energies.

\section{IV. Effective stress tensor and shape equation}
\label{app:StressTensor_ShapeEq}

The natural question that now arises is, Which shapes minimize the effective free energy~\eqref{eq:eff_Ham}? To answer this question, we now explore the field equations that govern the shape of the membrane as well as the stress tensor by implementation of the \textit{auxiliary variable method} introduced in~\cite{JemalGuven_2004}.

\subsection{First variation of the effective Hamiltonian by auxiliary variables}

The basic idea is to introduce Lagrange multipliers to impose the geometrical identities as holonomic constraints. In this procedure, an energy functional of the form $H\left[{\bf X}\right]=\int \dd A~\pazocal{H}\left[K_{ab}, g_{ab}, R\right]$ is replaced by a new functional {which includes all  relevant geometric restrictions}. In particular, this extended functional includes $\int \dd A\,{\bf f}^{a}\cdot\left({\bf e}_{a}-\partial_{a}{\bf X}\right)$, with the Lagrange multiplier ${\bf f}^{a}$ anchoring the tangent vector ${\bf e}_{a}$ to the embedding functions ${\bf X}$.
To enforce some of the definitions, we constrain our original Hamiltonian with Lagrange multipliers functions $\Lambda^{ab}$, ${\bf f}^{a}$, $\lambda_{n}$, $\lambda_{ab}$, and $\lambda_{\perp}^{a}$. This enlarged functional $H_c$ is defined as,
\begin{align*}
H_{c}& \coloneqq \int \dd A~\pazocal{H}\left[K_{ab}, g_{ab}, R\right]+\int \dd A ~{\bf f}^{a}\cdot\left({\bf e}_{a}-\partial_{a}{\bf X}\right)\nonumber\\
&+\int \dd A\left[\lambda_{\perp}^{a}\left({\bf e}_{a}\cdot{
\bf N}\right)+\lambda_{n}\left({\bf N}^2-1\right)\right]\nonumber\\
&+\int \dd A\left[\Lambda^{ab}\left(K_{ab}-{\bf e}_{a}\cdot\partial_{a}{\bf N}\right)+\lambda^{ab}\left(g_{ab}-{\bf e}_{a}\cdot{\bf e}_{b}\right)\right]\nonumber\\
&+\int \dd A~\Lambda_{R}\left[R-(g^{ab}K_{ab})^2+K_{ab}K_{cd}g^{ac}g^{db}\right],
\end{align*}
with ${\bf N}$ being the unit normal vector of the surface. We must now carry out variations with respect to each quantity. When performing a variation with respect to $K_{ab}$ or $g_{ab}$, it is convenient, to adopt the definitions
\begin{equation}
    \pazocal{H}^{ab}\coloneqq\frac{\partial \pazocal{H}}{\partial K_{ab}}\,, \, \pazocal{H}_{R} \coloneqq \frac{\partial \pazocal{H}}{\partial R}\,, \, T^{ab}\coloneqq -\frac{2}{\sqrt{g}}\frac{\partial (\sqrt{g}\pazocal{H})}{\partial g_{ab}}
\end{equation}
with $T^{ab}$ being the intrinsic stress tensor. In terms of these quantities, the variation with respect to $K_{ab}$ leads to $\Lambda^{ab}=-\pazocal{H}^{ab}-2\pazocal{H}_{R}(g^{ab}K-K^{ab})$, and that with respect to $g_{ab}$ leads to $\lambda^{ab}=\tfrac12 T^{ab} +2\pazocal{H}_{R}R^{ab}$. In the latter, the Gauss-Codazzi equation, $R_{ab}=K_{ab}K-K_{ac}K\indices{^{c}_{b}}$, is used.
 
Furthermore, ${\bf f}^{a}$, interpreted as the stress tensor of the membrane, is the N\"{o}ether current associated with the translational invariance of the membrane ${\bf X}\to {\bf X}+{\bf a}$ for any 3D constant vector ${\bf a}$; consequently, ${\bf f}^{a}$ satisfies the conservation law $\nabla_{a}{\bf f}^{a}=0$.

Importantly, variation with respect to the embedding functions ${\bf X}$ leads to the conservation of the stress tensor ${\bf f}^{a}$, i.e., $\nabla_{a}{\bf f}^{a}=0$ in the absence of external pressure fields; their inclusion is discussed in the following section. The stress tensor can be obtained from $\delta_{{\bf e}_{a}}H_c =0$, decomposing into ${\bf f}^{a}=f^{ab}{\bf e}_{b}+f^{a}{\bf N}$, where
\begin{equation}\label{eq:stress}
{\bf f}^{a}=\left(\Lambda^{ac}K_{c}^{b}+2\lambda^{ab}\right){\bf e}_{b}-\lambda^{a}_{\perp}{\bf N}.
\end{equation}
Finally, variation with respect to ${\bf N}$ results in
\begin{equation} \label{eq:var_wrtN}
\lambda^{a}_{\perp}{\bf e}_{a}+2\lambda_{n}{\bf N}+\nabla_{a}(\Lambda^{ab}{\bf e}_{a})=0.
\end{equation}
Using the Weingarten-Gauss equation, 
$\nabla_{a}{\bf e}_{b}=-K_{ab}\,{\bf N}$, we may rewrite~\eqref{eq:var_wrtN} as,
\begin{align}
\lambda_{\perp}^{a}&=-\nabla_{b}\Lambda^{ab},\nonumber\\
2\lambda_{n}&=\Lambda^{ab}K_{ab}\,.
\end{align}
Replacing into~\eqref{eq:stress}, this leads to the expression
 
\begin{align}
{\bf f}^{a}&=\left(T^{ab}-\pazocal{H}^{a}_{\hphantom{a}c} K^{cb}+R\pazocal{H}_{R}g^{ab}\right){\bf e}_{b}\nonumber \\
&\hphantom{=}\;-\left[\nabla_{b}\pazocal{H}^{ab}-2\left(\nabla_{b}\pazocal{H}_{R}\right)\left(g^{ab}K-K^{ab}\right)\right]{\bf N}.    
\end{align}
From the effective Hamiltonian~\eqref{eq:eff_Ham} one has $ \pazocal{H}_{R}=\kappa^{G}_{\rm eff}+\kappa^{{}_{(2)}}_{G}R$, $ \pazocal{H}^{ab}=\alpha g^{ab}K$, and the intrinsic stress tensor $    T^{ab}=\frac{\alpha}{2}K\left(4K^{ab}-Kg^{ab}\right)
    -\left(\sigma_{\rm eff}+\kappa^{\rm eff}_{G}R+\frac{1}{2}\kappa^{{}_{(2)}}_{G} R^2\right)g^{ab}$.  
Therefore, the tangent and normal components of the stress tensor are
\begin{align}
    \!\!f^{ab}&=\alpha K\bigg(K^{ab}-\frac{1}{2}g^{ab}K\bigg)-g^{ab}\bigg(\sigma_{\rm eff}
    -\frac{\kappa^{{}_{(2)}}_{G}}{2}R^2\bigg), \\
    f^{a}&=-\left[\alpha \nabla^{a}K+2\kappa^{{}_{(2)}}_{G}\nabla_{b}R\left(g^{ab}K-K^{ab}\right)\right],
\end{align}
respectively. Using the stress tensor, one can compute the force with which a piece $\mathcal{R}$ of the membrane acts on its surroundings. Indeed, according to the development in~\cite{DESERNO201511}, this force is given by $\int_{\partial\mathcal{R}}\dd s~\ell_{a}{\bf f}^{a}$, where $\dd s$ is the line element on the boundary $\partial\mathcal{R}$, and $\ell_{a}$ are the components of the unit normal vector outward $\partial\mathcal{R}$. 
The shape equation follows from the condition $\nabla_{a}f^{a}-K^{ab}f_{ab}=0${, that is}, 
\begin{eqnarray}
\label{eq:effshape_eq}
-\alpha\left[\Delta_{g}K+\frac{1}{2}K\left(K^2-2R\right)\right]+K_{ab}\mathcal{G}^{ab}=0, 
\end{eqnarray}
with $\mathcal{G}^{ab}$ defined as
\begin{align}
\mathcal{G}^{ab} \coloneqq \kappa_{G}^{{}_{(2)}}(&\nabla^{a}\nabla^{b}-g^{ab}\Delta_{g})R \nonumber \\
&\quad +\frac{1}{2}g^{ab}\bigg(\sigma_{\rm eff}-\frac{\kappa_{G}^{{}_{(2)}}}{2}R^2\bigg),\label{eq:gab_tensor}
\end{align}
where $\Delta_{g}$ is the Laplace-Beltrami operator acting on scalars compatible with the metric tensor $g_{ab}$.
Notice that $\kappa^{\rm eff}_{G}$ does not appear in the shape equation; this is a consequence of limiting our analysis to manifolds without boundaries, resulting in the $R$ term of the effective Hamiltonian being topological. Despite the non-linear nature of the shape equation~\eqref{eq:effshape_eq}, it is possible to deduce exact solutions corresponding to known structures constructed from graphene. 

On the one hand, $R=0$ and constant $K=1/r_{0}$ are a solution of~\eqref{eq:effshape_eq} which corresponds to a cylinder with radius $r_{0}=\sqrt{\alpha/(2 \sigma_{\rm eff})}$ when $\sigma_{\rm eff}>0$. Likewise, taking the conditions $K^{2}=2R$ and $K=2/r_{1}$ to be constant, one gets another solution of the Eq.~\eqref{eq:effshape_eq} which corresponds to a sphere with radius $r_{1}=(2\kappa_{G}^{{}_{(2)}}/ \sigma_{\rm eff})^{1/4}$ when $\sigma_{\rm eff}<0$. The condition $\sigma_{\rm eff}>0$ ($\sigma_{\rm eff}<0$) for a cylinder (sphere) imposes an upper (lower) bound on the equilibration temperature {$T<T_{\rm c}$} ($T>T_{\rm c}$), where $T_{\rm c}=\left(\frac{\sigma(\hbar v_{F})^{2}}{3g_{v}g_{s}\zeta(3){k_{B}^{3}}}\right)^{\frac{1}{3}}$. In particular, for the cylinder surface, the mean curvature and Ricci curvature satisfy 
\begin{eqnarray}
K_{\rm cyl}&=&\frac{1}{\ell_{T_{\rm c}}}\sqrt{\frac{6g_{v}g_{s}\zeta(3)k_{B}T_{\rm c}}{\alpha}}\left[1-\left(\frac{T}{T_{\rm c}}\right)^{3}\right]^{\frac{1}{2}},\nonumber\\
R_{\rm cyl}&=&0\label{cylsol},
\end{eqnarray}
whereas for the sphere surface, these curvatures satisfy
\begin{eqnarray}
K_{\rm sph}&=&\frac{2[1440\pi\zeta(3)]^{\frac{1}{4}}}{\ell_{T_{\rm c}}}\left(\frac{T}{T_{\rm c}}\right)^{\frac{1}{4}}\left[\left(\frac{T}{T_{\rm c}}\right)^{3}-1\right]^{\frac{1}{4}},\nonumber\\
R_{\rm sph}&=&\frac{1}{2}K^{2}_{\rm sph}.\label{sphsol}
\end{eqnarray}
Clearly, the temperature $T_{\rm c}$ distinguishes two separate phases where cylindrical and spherical surface formation occurs. This distinction is a consequence of the contrast dependence of the effective elastic coefficients with respect to the thermal wavelength; indeed, while the surface tension coefficient behaves as $\beta\delta\sigma_{\rm eff}\sim \ell_{T}^{-2}$, the quantum elastic coefficient behaves as $\beta \kappa_{G}^{{}_{(2)}}\sim \ell_{T}^{2}$. This is another way to see that there are two opposite temperature regimes for the formation of the surfaces. For convenience, we call them the high- and low-temperature regimes; these regimes can be defined as $T\gg T_{\rm c}$ and $T\ll T_{\rm c}$, respectively, where $T_{\rm c}$ is the same characteristic temperature defined above. In the low-temperature regime, the density energy $\frac{1}{2}\beta \kappa_{G}^{{}_{(2)}}R^{2}$ dominates the elastic behavior, while in the high-temperature region, the term $\delta\sigma_{\rm eff}$ dominates the behavior. 
In the low-temperature regime, the shape equation reduces to $K_{ab}\left[\nabla^{a}\nabla^{b}-g^{ab}\left(\Delta_{g}+\frac{1}{4}R\right)\right]R=0$.
Note that this equation is identically satisfied for $R=0$. Thus, all surfaces with $R=0$ represent solutions in this regime, encompassing various types of geometric configurations known as \textit{developable surfaces}; generated by sweeping a straight line in space and revolving around an axis: planes,  cylinders, conical surfaces, tangent surfaces, and a combination of pieces of them~\cite{DoCarmo}. In addition, in the high-temperature regime, the shape equation reduces to $\delta\sigma_{\rm eff}K=0$ or $K=0$, which corresponds to a plethora of structures known as {\it minimal surfaces}~\cite{DoCarmo}.

\section{V. Stability analysis of the geometric configurations}
\label{sec:Stability_of_geoconf}

\subsection{Second variation of the effective Hamiltonian}
Following~\cite{capovilla2003deformations}, the second variation of a surface functional invariant under reparametrizations $H[{\bf X}]=\int \dd^{2}x\sqrt{g} f({\bf X})$ can always be expressed in the form 
\begin{equation}\label{eq:second_var}
    \delta^{2}H=\int \dd A\, \Phi\mathcal{L}_{f}\Phi,
\end{equation}
for some local differential operator $\mathcal{L}_{f}$, where the change in the embedding function is given by ${\bf X}\to{\bf X}+\delta{\bf X}$, with $\delta{\bf X}=\Phi{\bf n}+\Phi^{a}{\bf e}_{a}$ where $\Phi$ and $\Phi^a$ are smooth, scalar and vector fields. In addition, it has been assumed that the surfaces are closed without boundaries. In particular, the operator $\mathcal{L}_f$ can be obtained by performing the normal variation $\mathcal{L}_f \Phi=\delta_{\perp}[\sqrt{g}\mathcal{E}(f)]$, where $\mathcal{E}(f)$ is the factor obtained in the first normal variation of the functional. In particular, the Euler-Lagrange equations are at equilibrium  $\mathcal{E}(f)=0$. Here, we focus on the second-order variation of the functional $\tfrac{1}{2}\kappa_{G}^{{}_{(2)}}\int \dd A\, R^{2}$ because the variation of the other terms in the effective Hamiltonian was already computed in~\cite{capovilla2003deformations}. After a tedious but straightforward calculation,
\begin{widetext}
\begin{align*}
\frac{1}{4}\mathcal{L}_{R^{2}}&= K\left[K^{ab}(\nabla_{a}\nabla_{b}R)-K(\Delta_{g}R)-\frac{1}{4}R^{2}K\right]+(\nabla_{a}\nabla_{b}R)\left[-\nabla^{a}\nabla^{b}+(KK^{ab}-R^{ab})-4K^{ac}K\indices{_{c}^{b}}\right]\nonumber\\
&+(K^{ab}-Kg^{ab})\nabla_{a}\nabla_{b}\left\{2\nabla_{c}\left[(K^{cd}-g^{cd}K)\nabla_{d}\left(\cdot\right)\right]-RK\left(\cdot\right)\right\}-\nabla_{c}R\left[K_{a}^{c}\nabla_{b}+K_{b}^{c}\nabla_{a}-K_{ab}\nabla^{c}+(\nabla_{a}K_{b}^{c})\right]\nonumber\\
&-(\Delta_{g}R)\left[-\Delta_{g}+(R-K^{2})\right]+2KK^{ab}\nabla_{a}\nabla_{b}R
+K\nabla_{a}K\nabla^{a}R+2KK^{ab}\nabla_{a}R\nabla_{b}-K^{2}\nabla^{a}R\nabla_{a}\nonumber\\&-\frac{1}{2}RK\left\{2\nabla_{c}\left[(K^{cd}-g^{cd}K)\nabla_{d}\left(\cdot\right)\right]\right\}
-\frac{1}{4}R^{2}\left[-\Delta_{g}+(R-K^{2})\right]\,.
\end{align*}
\end{widetext}
As in~\cite{capovilla2003deformations}, we have the following expressions for the operators $\mathcal{L}_{K^2}$ and $\mathcal{L}_{1}$, defined in~\eqref{eq:second_var}: 
\begin{align*}
    \mathcal{L}_{K^{2}}&= 2\Delta_{g}^{2}+(K^{2}-4R)\Delta_{g}+4KK^{ab}\nabla_{a}\nabla_{a}+2R^2\nonumber\\
    &\hphantom{=}\, + 2K^4 - 5RK^2 + 12K^{ab}\nabla_{a}\nabla_{b}K-K\Delta_{g}K \nonumber\\
    &\hphantom{=}\, +(\nabla^{a}K)(\nabla_{a}K),
\end{align*}
and, for $f=1$, $\mathcal{L}_{1}=-\Delta_{g}+R$.

\subsection{Stability analysis of the cylinder}

In the case of a cylinder one parametrization is ${\bf X}(\theta, z)=(r_{0}\cos\theta, r_{0}\sin\theta, z)$, where we recall that $r_{0}=\sqrt{{\alpha}/{2\sigma_{\rm eff}}}$ is the radius of the cylinder with $\sigma_{\rm eff}>0$. Under this parametrization, the cylinder line element is $\dd s^2 = r^2 \dd \theta^2 + \dd z^2$. This
results in a single nonvanishing component of the extrinsic curvature $K_{\theta\theta}=r_{0}$ and inverse $K^{\theta\theta}=1/r^{3}_{0}$.
Additionally, the mean curvature $K=1/r$, and the Ricci curvature $R=0$. The Laplace-Beltrami operator on scalars is $\Delta_{\rm cyl}=r^{-2}_{0}\partial^{2}_{\theta}+\partial^{2}_{z}$. The spectrum of this operator is $-\frac{1}{r^{2}_{0}}[m^2+(kr_{0})^{2}]$, for $m\in\mathbb{Z}$ and $k\in\mathbb{R}$. Using $\mathcal{L}_{K^2}, \mathcal{L}_{1}$, and $\mathcal{L}_{R^{2}}$, we can compute the spectrum of the operator $\mathcal{L}_{\rm cyl}$ for the energy density of the effective Hamiltonian,
\begin{align}
    {\rm spec}(\mathcal{L}_{\rm cyl}) &= \frac{2\sigma_{\rm eff}}{r^{2}_{0}}\left\{m^{4}+2[(kr_{0})^2-1]m^2+1\right\}\nonumber \\
    &\hphantom{=}\; + \frac{\left(2\sigma_{\rm eff}\right)^{2}}{\alpha}\left[1+\frac{8\kappa^{{}_{(2)}}_{G}\sigma_{\rm eff}}{\alpha^{2}}\right](kr_{0})^{4}\,.
\end{align}
Observe that the first term is always positive, whereas the second term is positive only if the condition $1+8\alpha^{-2}\kappa^{{}_{(2)}}_{G}\sigma_{\rm eff}>0$ holds. This implies the following inequality for the temperature 
\begin{equation}\label{eq:ccyl}
\frac{T_{*}}{1-T^{3}_{*}}>c \coloneqq c_{\rm cyl}\,,
\end{equation}
where we have defined the reduced temperature $T_{*}=T/T_{c}$ and $c=8\alpha^{-2}g_{v}g_{s}\ell^{2}_{T_{c}}\sigma k_{B}T_{c}$. The minimum value of the temperature where the cylinder configuration is stable is given by 
\begin{equation}\label{eq:Tmin/Tc}
   \frac{T_{\rm min}}{T_{c}}=\left(\frac{2\lambda}{3}\right)^{\frac{1}{3}} \left[\frac{1}{2c}\left(\frac{2}{3}\right)^{\frac{1}{3}}-\lambda^{-\tfrac{2}{3}}\right],
\end{equation}
with $\lambda=9 c^3 +\sqrt{3c^3 (27 c^3+4)}$ Additionally, one can show that the image of the right-hand side of Eq.~\eqref{eq:Tmin/Tc} is the interval $[0,1]$ for any value of $c\in \mathbb{R}$.

\subsection{Stability analysis of the sphere}

In contrast, the sphere does not require explicit parametrization. It suffices to
use the fact that its extrinsic curvature tensor satisfies $K_{ab}=\frac{K}{2}g_{ab}$ and $K^2=2R$. Let us recall that the sphere has a radius $r_{1}=(2\kappa^{{}_{(2)}}_{G}/\sigma_{\rm eff})^{1/4}$, or, equivalently, Ricci curvature $R=(2\sigma_{\rm eff}/\kappa^{{}_{(2)}}_{G})^{1/2}$, with $\sigma_{\rm eff}<0$. Also, we take advantage of the fact that the Laplace-Beltrami operator on scalars for the sphere is written in terms of the angular momentum $\hat{\bf L}$ as $\Delta_{S^{2}}=-\frac{R}{2}\hat{\bf L}^{2}$, whose eigenvalues are given by $\ell(\ell+1)$ for $\ell\in \mathbb{N}\cup \{0\}$. The spectrum of the operator $\mathcal{L}_{S^2}$ for the energy density of the effective Hamiltonian is obtained using $\mathcal{L}_{K^2}, \mathcal{L}_{1}$, and $\mathcal{L}_{R^{2}}$:
\begin{equation}
    {\rm spec}(\mathcal{L}_{S^{2}})=\frac{1}{4}\alpha R^2 f^{{}_{(1)}}_{\ell}+\frac{1}{2}\sigma_{\rm eff}Rf^{{}_{(2)}}_{ \ell},
\end{equation}
where
\begin{align*}
f^{{}_{(1)}}_{\ell}&=\ell(\ell+1)\left[\ell(\ell+1)-2\right],\nonumber \\
f^{{}_{(2)}}_{\ell}&=2 \left[\ell(\ell+1)-2\right]\left[\ell(\ell+1)-\tfrac{5}{2}\right]+\ell(\ell+1)+1.
\end{align*}
Note that the sequence $f^{{}_{(1)}}_{\ell}\geq 0$, whereas the sequence $f^{{}_{(2)}}_{\ell}>0$ for all $\ell\in \mathbb{N}\cup \{0\}$. The stability condition should satisfy ${\rm spec}(\mathcal{L}_{S^{2}})>0$ for all $\ell$, implying $-2\sigma_{\rm eff}/(\alpha R)<f^{{}_{(1)}}_{\ell}/f^{{}_{(2)}}_{\ell}$. The ratio of the sequences is then bounded, $f^{{}_{(1)}}_{\ell}/f^{{}_{(2)}}_{\ell}\leq 24/35$. Substituting the expression for the Ricci curvature in terms of the temperature by a straightforward calculation, we obtain the condition
\begin{equation}\label{eq:cS2}
    \frac{T_{*}^3-1}{T_{*}}<\left(\frac{24}{35}\right)^{2}\frac{\alpha^2 1440\pi \zeta(3)}{\sigma^{2}\ell^{4}_{T_{c}}}=d \coloneqq c_{S^{2}}\,.
\end{equation}
Recall that $T_{*}\geq 1$ is the condition found for the sphere solution. Thus, the maximum value of $T_{*}$ that satisfies the above inequality is the real root of the third-order polynomial $P(x)=x^{3}-dx-1$ for $x\geq 1$.

\subsection{Stability analysis of the minimal surfaces}

Minimal surfaces are characterized by $K=0$; we make use of the operators $\mathcal{L}_{K^{2}}$, and $\mathcal{L}_{1}$ to show that 
\begin{equation*}
\mathcal{L}_{\rm MinS}=\alpha(-\Delta_{g}+R)^{2}+\sigma_{\rm eff}(-\Delta_{g}+R), 
\end{equation*}
which certainly will give a positive spectrum. Note that $\alpha>0$ is required for stability. As a consequence, at high temperatures the contribution from $\sigma_{\rm eff}$ dominates resulting in ${\rm spec}(\mathcal{L}_{\rm MinS})<0$, i.e., unstable structures.

\subsection{Stability analysis of the developable surfaces}

For developable surfaces, $R=0$, we {make use of} the operator $\mathcal{L}_{R^{2}}$; we remind the reader that $R^2$ sector of the free energy dominates at low temperatures. Using the Codazzi-Mainardi equation $\nabla_{a}K^{ab}=\nabla^{b}K$, we find 
\begin{equation*}
    \mathcal{L}_{\rm DevS}=4\kappa^{{}_{(2)}}_{G}\mathcal{O}^{2},
\end{equation*}
where $\mathcal{O}=(K^{ab}-Kg^{ab})\nabla_{a}\nabla_{b}$. Since $\kappa^{{}_{(2)}}_{G}$ is negative, the developable surfaces are automatically unstable.

\section{VI. Phase diagram of 2D graphitic materials}

\begin{figure}
    \includegraphics[width=\linewidth]{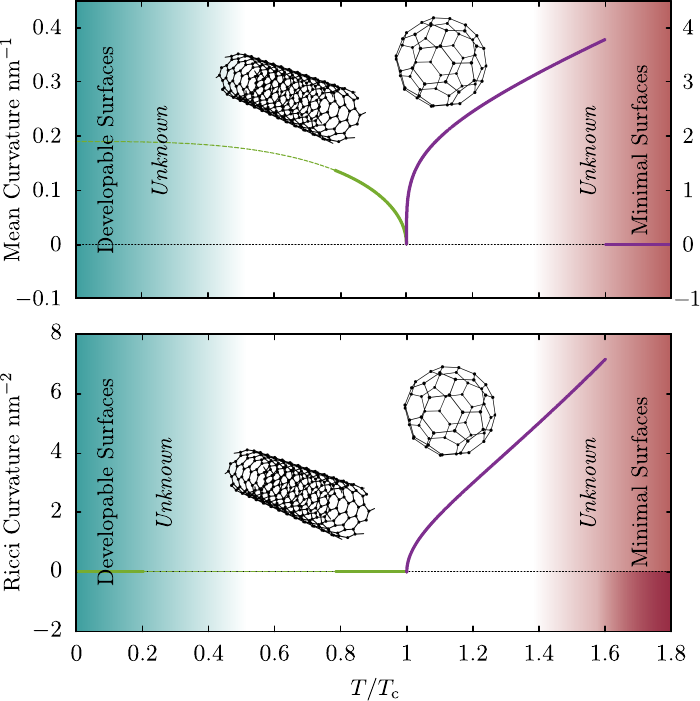}
    \caption{Mean and Ricci curvatures as a function of temperature. The green (purple) lines denote the values for cylindrical (spherical) structures in the nanometer scale. The solid lines shows the stability region starting at $T_{\rm min}= 751.287\,$K for carbon nanotubes. The critical temperature $T_{\rm c} = 957.757\,$K is obtained by imposing the constraints $r_{C60}<1.3\,$nm and $r_{\rm nt}<7\,$nm at synthesis temperatures $T_{C60}=1000\,$K and $T_{\rm nt}=720\,$K within the range of known values in the literature~\cite{Sugai2000}. The values of $T_{\rm min}$ and $T_{\rm c}$ were obtained using $\alpha=1.44~{\rm eV}$~\cite{YujieWei}, and Fermi velocity $v_F = 0.85 \times 10^{6} \,{\rm m/s}$~\cite{hwang2012fermi} (see Appendix~D for details).}
    \label{fig:phase_diagram}
\end{figure}

A phase diagram for the conformation of surfaces above is depicted  
in Fig.~\ref{fig:phase_diagram} through the mean and Ricci curvatures versus the reduced temperature $T/T_{\rm c}$. The phase diagram shows six regions; from left to right, they are a developable surface phase for  $T\ll T_{\rm c}$, a cylinder phase for $T<T_{\rm c}$, a sphere phase for $T_{\rm c}<T$, a minimal surface phase for $T\gg T_{\rm c}$ and two unknown regions. Note that from the solutions~\eqref{cylsol} and~\eqref{sphsol} we cannot infer a lower value of temperature for the cylinder or an upper value of temperature for the sphere phase. However, we can infer these temperatures from the mechanical stability analysis in Sec.~V.

Indeed, the second variation of the effective Hamiltonian~\eqref{eq:eff_Ham} reveals that cylindric structures are stable only at $T>c_{\rm cyl}T_{\rm c}[1-({T/T_{\rm c}})^{3}]$ and spheres are stable only at $T>c_{S^{2}}T_{\rm c}[(T/T_{\rm c})^{3}-1]$, with the constants $c_{\rm cyl}$ and $c_{S^{2}}$ from~\eqref{eq:ccyl} and~\eqref{eq:cS2}, respectively. The large value of $T_{\rm max}$ obtained from the last inequality implies the stability of the spherical structures to all practical values of temperature. In addition, the minimal surface solutions under the Hamiltonian $\int \delta\sigma_{\rm eff}dA$ are unstable; however, if we do not neglect the bending energy $\frac{\alpha}{2}\int \dd A K^{2}$, the shape equation reduces to $-\alpha\left[\Delta_{g}K+\frac{1}{2}K\left(K^2-2R\right)\right]+\sigma_{\rm eff}K=0$, where the minimal surfaces, $K=0$, are still solutions but now the surfaces are stable. Therefore, the stable minimal surfaces are found in the high-temperature regime while neglecting the $\kappa^{{}_{(2)}}_{G}$ terms but not the bending coefficient $\alpha$. In contrast, the developable surface of graphene dominated by the term $\frac{1}{2}\kappa^{{}_{(2)}}_{G}R^{2}$ appears to be unstable under mechanical deformation. This instability may be related to the tendency of the membrane to crumple~\cite{PhysRevLett.120.261601, Castro2024}. 

\section{VII. Configurations with external pressure field}

The shapes described by~\eqref{eq:effshape_eq} are believed to be the outcome of the mechanical equilibrium process during the final stage of the material synthesis---spontaneously formed, without the influence of external forces. However, membranes are often subjected to external fields. To account for this, we add a source term, $-\int dA~{\bf J}\cdot{\bf X}$, to the effective Hamiltonian~\eqref{eq:eff_Ham}. The shape equation for the membrane with source terms now requires ${\bf J} = \delta_{{\bf X}} H_{\rm eff}$; thus, the modified shape equation is then given by $\nabla_{a}{\bf f}^{a}={\bf J}$, where ${\bf J}$ is the external field acting on each point of the membrane. 

Confirmed via atomic force microscopy~\cite{Ashino1, *Ashino2}, the classical elasticity of a piece of a graphene membrane is describable by the Helfrich Hamiltonian $H_{\rm Helfrich}[{\bf X}]=\int dA\left[\frac{\alpha}{2}K^2+\kappa_{G}R\right]$. A possible choice of the source term is ${\bf J}=\nabla_{a}{\bf f}^{a}_{\rm Helfrich}$ where 
\begin{equation}
{\bf f}^{a}_{\rm Helfrich}=\alpha K\bigg(K^{ab}-\frac{1}{2}g^{ab}K\bigg){\bf e}_{b}-\alpha\nabla^{a}K~{\bf N},
\label{Helfrich-StressTensor}
\end{equation}
is the stress tensor of the Helfrich Hamiltonian since one may then be interested in configurations that result from balancing out classical elastic forces, i.e., equating the internal stresses to the external source. In this situation, the Helfrich internal stresses of the graphene membrane~\cite{Ashino1, *Ashino2} are balanced out; therefore, the shape equation can be succinctly written as $K_{ab}\mathcal{G}^{ab}=0$, with $\mathcal{G}_{ab}$ defined in~\eqref{eq:gab_tensor}.

Let us remark that the field equation $\mathcal{G}_{ab}=0$ gives a shape equation for all surfaces that extremize the functional $\int \dd A \left[\sigma_{\rm eff}+\kappa_{G}^{ \rm eff}R+\frac{1}{2}\kappa^{{}_{(2)}}_{G}R^2\right]$ with respect to changes in the metric tensor $g_{ab}$. In particular, surfaces of constant Ricci curvature $R=\pm R_{\rm sph}$, with $R_{\rm sph}$ defined in~\eqref{sphsol}, are solutions of $\mathcal{G}_{ab}=0$. The positive sign corresponds to a sphere, whereas the negative sign yields a Beltrami pseudosphere. Indeed, this surface has been found to be energetically stable in numerical simulations~\cite{Taioli2016} and 
is considered a candidate to reproduce the Hawking-Unruh effect in graphene systems~\cite{IORIO2012, Iorio2014}.

\subsection{Pressure field in the Beltrami pseudosphere}

The extrinsic curvature tensor of a sphere satisfies $K_{ab}=\frac{1}{2}K g_{ab}$, with constant $K$ implying that $\nabla_{a} {\bf f}^{a}_{\rm Helfrich}=0$; hence, no pressure field is required to produce spherical surface. In contrast, a Beltrami pseudosphere cannot be formed in their absence. The force associated with this pressure field over {an} area section $\mathcal{R}$, can be computed as ${\bf F}=\int_{\mathcal{R} } dA\,{\bf J}$. Because ${\bf J}=\nabla_{a}{\bf f}^{a}_{\rm Helfrich}$ one can use the Stokes theorem to rewrite the integral as ${\bf F}=\oint_{\gamma}ds~\ell_{a}{\bf f}^{a}_{\rm Helfrich}$, where $\gamma=\partial\mathcal{R}$ is the curve which encloses the region $\mathcal{R}$. We can directly compute ${\bf J}$ from the covariant derivative of the Helfrich stress tensor~\footnote{Extensive analysis of these forces is beyond the scope of this paper and will be the subject of subsequent work to be reported elsewhere}, 
\begin{eqnarray}
{\bf J}=-\alpha\left[\Delta_{g}K+\frac{1}{2}K\left(K^2-2R\right)\right]{\bf N},
\end{eqnarray}
where ${\bf N}$ is the normal vector of the surface. To proceed, we use the following parametrization of the Beltrami pseudosphere: 
\begin{align}
{\bf X}(\varphi, u)&=\left(re^{\frac{u}{r}}\cos\varphi, re^{\frac{u}{r}}\sin\varphi, \right.\nonumber \\
&\hphantom{=}\qquad\; \left. r\left[{\rm arctanh} f(u)-f(u)\right]\right),
\end{align}
with $f(u)=\sqrt{1-e^{2u/r}}$, where $u\in \left(-\infty, 0\right]$ and $\varphi\in [0, 2\pi)$. Here $u=0$ maps to the maximal circle of radius $r$ at the bell of the trumpet, a singular boundary known as the {\it Hilbert horizon}~\cite{Taioli2016}, and $u\to-\infty$ is the mouthpiece of the trumpet.

The tangent vectors of this surface are given by ${\bf e}_{\varphi}=re^{\frac{u}{r}}\boldsymbol{\ell}$ and ${\bf e}_{u}=e^{\frac{u}{r}}{\bf t}-f(u)\hat{\bf z}$, with unit vectors $\boldsymbol{\ell}=(-\sin\varphi, \cos\varphi, 0)$, ${\bf t}=\left(\cos\varphi, \sin\varphi, 0\right)$, and $\hat{\bf z}=\left(0,0,1\right)$ being an orthonormal set. Using these tangent vectors, the line element of the Beltrami trumpet is given by $\dd s^2=\dd u^2+R^2(u)\dd \varphi^2$ with $R(u)=re^{u/r}$.

In addition, the normal vector can easily be computed as ${\bf N}=f(u){\bf t}+e^{\frac{u}{r}}\hat{\bf z}$; differentiation of the normal vector with respect to $u$ and $\varphi$ yields ${\bf N}_{u}=\frac{1}{r}\left(f(u)-f(u)^{-1}\right){\bf t}+\frac{1}{r}e^{\frac{u}{r}}\hat{\bf z}$, and ${\bf N}_{\varphi}=f(u)\boldsymbol{\ell}$, respectively. 
We can now compute the components of the extrinsic curvature tensor $K_{ab}$: $K_{\varphi\varphi}={\bf e}_{\varphi}\cdot{\bf N}_{\varphi}=re^{\frac{u}{r}}f(u)$, $K_{uu}={\bf e}_{u}\cdot{\bf N}_{u}=-\frac{1}{r}e^{\frac{u}{r}}f(u)^{-1}$, and $K_{u\varphi}=K_{\varphi u}=0$. The components of $K^{ab}=g^{ac}g^{bd}K_{cd}$, read $K^{uu}=-\frac{1}{r}e^{\frac{u}{r}}f(u)^{-1}$ and $K^{\varphi\varphi}=\frac{1}{r^{3}}e^{-3\frac{u}{r}}f(u)$. The mean curvature is thus
\begin{equation}
K=g^{ab}K_{ab}=-\frac{1}{r}e^{\frac{u}{r}}f(u)^{-1}+\frac{1}{r}e^{-\frac{u}{r}}f(u),
\end{equation}
where one can identify the principal curvatures $k_{1}=-\frac{1}{r}e^{\frac{u}{r}}f(u)^{-1}$ and $k_{2}=\frac{1}{r}e^{-\frac{u}{r}}f(u)$.
The Laplace-Beltrami operator for the Beltrami trumpet is 
$\Delta_{g}\cdot =R^{-1}(u)\partial_{u}\left(R(u)\partial_{u}\cdot\right)+R^{-2}(u)\partial^{2}_{\varphi}\cdot$. $K$ depends only on $u$; hence, only the first term remains. By straightforward calculation, the pressure field is given by 
\begin{align}
{\bf J} =-\frac{\alpha}{2r^3}\frac{e^{\frac{3u}{r}}}{f^{3}(u)}&\left[\left( e^{-\frac{2u}{r}}f^{2}(u)-1\right)\left(e^{-\frac{4u}{r}}f^{4}(u)+1\right)\right.\nonumber \\
&\quad\left.-4e^{-2\frac{u}{r}}\right]{\bf N}.
\end{align}
The pressure field ${\bf J}$ near the tip is ${\bf J}\approx -({\alpha}/{2r^3})\exp\left(-{3u}/{r}\right){\bf t}$, supporting the trumpet's mouthpiece, whereas near the bell of the trumpet, one has the asymptotic behavior ${\bf J}\approx ({5\sqrt{2}\alpha}/{8r^3} )\left(-{u}/{r}\right)^{-{3}/{2}}\hat{\bf z}$. Here, recall that the unit vectors are given by ${\bf t}=(\cos\varphi, \sin\varphi, 0)$ and $\hat{\bf z}=(0,0,1)$. This means that there is a compression near the tip and tension near the bell trumpet that lift the membrane, while it is held fixed at the Hilbert horizon. Our results are consistent with the qualitative analysis in~\cite{Taioli2016}.

\section{VIII. Conclusions}
In this paper, we provided a phase diagram for the shapes that a curved sheet of graphene can have, under the assumption that the electronic degrees of freedom are described by the Dirac curved model.
It is noteworthy that despite the simple and naive model for curved graphene used here, it captures most of the equilibrium geometries that have been experimentally observed for graphene~\cite{TERRONES2010351}. At low temperatures $T \ll T_{\rm c}$ one encounters the family of developable surfaces, with $R=0$ indicating where nanocones and nanoribbons can be expected and cylindrical structures such as single-wall carbon nanotubes (CNTs) being predicted for $T_{\rm min}<T< T_{\rm c}$.
At $T > T_{\rm c}$, a family of spheres represented by fullerenes such as buckminsterfullerene $C_{60}$ occurs, and at $T \gg T_{\rm c}$ elusive minimal surfaces with $K=0$, with schwarzites, triply periodic minimal surfaces~\cite{Terrones}, and glasslike Carbon~\cite{nano11071694} being among the candidates, occur. 

In our approach, the negative Casimir-like correction $\delta\sigma_{\rm eff}$ from chiral fermions imposes a tight constraint on the classical contribution to the surface tension $\sigma$, implying the critical temperature $T_{\rm c}$ to be measured in the laboratory.
Let us remark that Eqs.~\eqref{cylsol} and~\eqref{sphsol} indicate that the cylindrical and spherical phases are mutually exclusive and that carbon nanotubes are produced at a lower temperature than the temperature required for fullerene formation. Stability analysis of these structures reveals the existence of minimum and maximum temperature values required for the production of carbon nanotubes and fullerenes, respectively. These results are in qualitative agreement with findings in the literature because the formation rate of CNTs, via thermal pyrolysis synthesis~\cite{SRISUMA20211373}, increases above $700\,$K and decreases after $1000\,$K. Also, as shown in~\cite{Sugai2000}, fullerenes $C_{60}$ and $C_{70}$ do not form at $298.15\,$K but are effectively produced above $1073.15\,$K. Our model does not specify an exact value for the critical temperature, but we can use experimental data to establish a precise constraint of $T_{\rm c} = 957.757\,$K, which allowed us to draw the phase diagram in Fig.~\ref{fig:phase_diagram}. Although negative Gaussian curvature has not been produced in the laboratory or found in nature, there is a reasonable expectation to do so. The Beltrami trumpet-shaped membrane with constant $R<0$~\cite{IORIO2012, Iorio2014} backed by numerical simulations~\cite{Taioli2016} has proven to be a promising candidate. Here, we showed that considering the Dirac field thermal fluctuations in the membrane dynamics reveals that the surface is possible once the internal Helfrich stresses are accounted for. Importantly, it may be necessary to seek the synthesis at yet higher temperatures than originally thought~(Fig.~\ref{fig:phase_diagram}). 

As a matter of perspectives, it may be interesting to explore what other \textit{class} of solutions, i.e. expected surfaces, Eq.~\eqref{eq:effshape_eq} allows. However, as evidenced via Monge gauge, the shape equation often results in a complicated nonlinear fourth-order differential equation for the height function in which appropriate boundary conditions and approximations should be carefully considered. The consistency of~\eqref{eq:effshape_eq} with the currently observed structures predicts a strict bound on the values for the surface tension of the graphene membrane. However, under a conformal transformation, the shape equation constrains the conformal factor, resulting in a more tractable second-order partial differential equation to explore nontrivial deformations of simple geometries. Additionally, the inclusion of an electromagnetic field would provide a way to manipulate the elasticity of the membranes since the electromagnetic fields are in the ambient space where the graphene membrane is embedded and may correct the elastic bending parameter. Furthermore, one can consider a general non-abelian gauge field to control the valley's degrees of freedom. In this case, the heat-kernel expansion of the dynamics would result in terms involving the gauge field strength tensor coupled to the geometry. This could lead to non-Abelian Wilson lines as solutions, effectively explaining topological defects; for instance, the magnetic monopole in the case of the sphere~\cite{vozmediano2010gauge}. Regardless,
to accurately understand the carbon shapes, a low-energy field theory developed from an adequate curved tight-binding description is required; a discussion in this direction can be found in~\cite{MattWiseman1,MattWiseman2,CommentPaper,ReplyToComment}.

\section{Acknowledgments}
P.A.M. would like to thank R. Kanai for supporting this research. P.C.-V. would like to acknowledge the Autonomous University of Chiapas for its continued support and commitment to advancing research and education. Furthermore, P.C.-V. recognizes the financial assistance provided by SNI-CONAHCyT (No. CVU 92896).

\section{Appendix}
\subsection{A. Dirac Hamiltonian description of the graphene monolayer}
\label{app:Dirac_description_graph}

Low-energy excitations with momentum near any of the Dirac points $K_{\pm}$ in crystalline structures such as graphene exhibit a linear dispersion relation which can be effectively described by a continuous model that reduces to the (1+2)-dimensional Dirac equation~\cite{cortijo2007electronic}. Indeed, the scale that governs the Dirac particle gas confined to the flat sheet, the Fermi velocity $v_F$, is roughly a hundredth of the speed of light. This brings us to a ``relativistic" behavior in which the particle-hole system is describable as chiral fermions.
For each Dirac point, the components of the two-spinor $\psi = (\varphi_A , \varphi_B)$ describe the wave function of sublattices $A$ and $B$, respectively. When considering the honeycomb lattices, one must accommodate states from these two-spinors that account for both $K_{\pm}$. These can be combined into a four component Dirac spinors $\Psi = (\psi_+ , \psi_- )$, whose Dirac action we cast as the effective Hamiltonian,
\begin{equation}
    H = \int \dd^2 x \sqrt{g} \Psi^{\dagger}
    \begin{pmatrix}
    \pazocal{H}_{K_{+}} & 0 \\
    0 & \pazocal{H}_{K_{-}}
    \end{pmatrix}
    \Psi,
\end{equation}
with $\pazocal{H}_{K_{\pm}} =-i\hbar v_{F}\gamma^{0}\underline{\gamma}^{a}\nabla_{a}$ corresponding to the Hamiltonian density near $K_{\pm}$. From $\pazocal{H}_{K_{\pm}}$ we can see that the induced effective Fermi velocity of the charge carriers in induced $v^{{\rm eff}, \ell}_{a}(x)=v_{F}e^{\ell}_{a}(x)$ acquires space-time dependence from the vielbein in the curved gamma matrices. The vielbein may be locally expanded via Riemann normal coordinates $e^{\ell}_{a}(y)=\delta^{\ell}_{a}+\frac{1}{3}R^{\ell}_{pqa}(0)y^{p}y^{q}+\cdots$, where $R^{\ell}_{pqa}(0)$ are the components of the Riemann curvature tensor evaluated at a fiducial point~\cite{Muller1999}. The above field theoretic Hamiltonian corresponds to the action of two identical Dirac fields defined using a $2+1$ ultrastatic space-time metric. In the flat limit, the Fermi velocity reduces to $v_F =3ta/2$, where $t$ is the nearest-neighbor hopping parameter and $a$ is the lattice distance, being $a = 1.42\,$\r{A}. In particular, graphene may be described in this way, i.e., as quasiparticles subject to a curved space-time, as a product of the $\sigma$ bonds formed by the carbon atoms in the monolayer allowing the membrane to withstand elastic strains. The accumulation of topological defects such as pentagon or heptagon dislocations result into ripples modeled by the metric $g_{ab}$. For most cases it is often enough to consider one Dirac point while accounting for the valley degeneracy $g_v = 2$. In this way, the memory of the graphene lattice structure is carried through the scale $v_F$, and $g_s g_v = 4$ constants at the effective action and curvature values.

\subsection{B. Notation and Fourier decomposition of the Dirac Field}
\label{app:Fourier_Dirac}

Let us perform a decomposition in Fourier modes as follows
\begin{eqnarray}
    \Psi\left(\tau, x\right)&=&\frac{1}{\sqrt{\beta}}\sum_{n\in\mathbb{Z}}\psi_{n}\left(x\right)e^{-i\omega_{n}\tau}, \\
    \psi_{n}(x)&=&\frac{1}{\sqrt{\beta}}\int_{0}^{\beta}\dd\tau\Psi(\tau, x)e^{i\omega_{n}\tau},
\end{eqnarray}
where $\omega_{n}=(2n+1)\pi/\beta$, with $n\in\mathbb{Z}$, are the Matsubara frequencies. The orthogonality relation for the Fourier basis is $\int_{0}^{\beta}d\tau e^{i\tau\left(\omega_{n}-\omega_{n^{\prime}}\right)}=\beta\delta_{nn^{\prime}}$, and the completeness relation $\sum_{n\in\mathbb{Z}}e^{i\omega_{n}\left(\tau-\tau^{\prime}\right)}=\beta \delta(\tau-\tau^{\prime})$. Using the Fourier decomposition of the Dirac field, and the orthogonality relation, the action adopts the following frequency representation
\begin{eqnarray}
    S(\Psi^{\dagger}, \Psi)=\sum_{n\in\mathbb{Z}}\int_{\Sigma}\dd^{2}x\sqrt{g}~\psi^{\dagger}_{n}\left(x\right)\left(-i\omega_{n}+\hat{\mathcal{H}}\right)\psi_{n}(x).\nonumber
\end{eqnarray}
The functional measure in the Fourier basis is written as $\mathcal{D}\left(\Psi^{\dagger}, \Psi\right)=\prod_{n\in\mathbb{Z}}\mathcal{D}\left(\psi^{\dagger}_{n},\psi_{n}\right)$, thus the functional integral adopts the following expression
\begin{equation}
    Z=\prod_{n\in\mathbb{Z}}\left[\int \mathcal{D}\bar{\psi}_{n}\mathcal{D}\psi_{n} e^{-\int_{\Sigma}\dd^{2}x\sqrt{g}~\bar{\psi}_{n}\left(x\right)\left(i\omega_{n}\gamma^{0}+\slashed{D}_{E}\right)\psi_{n}(x)}\right]\nonumber
    \label{partitionfunction3}
\end{equation}
where $\slashed{D}_{E}=\gamma^{0}\hat{\mathcal{H}}=i\underline{\gamma}^{a}\nabla_{a}$ is a Euclidean Dirac operator with $\{\psi_{n}(x)\}$ being Grassmann variables. These integrals can be carried out directly, leading to~\eqref{eq:finTfermZ}.

\subsection{C. Heat kernel expansion of the Dirac Operator}
\label{app:Heat-Kernel_Expansion}

Let us make a few remarks with respect to the local expansion of the heat kernel~\eqref{eq:kernel_exp}. As described~\cite{VASSILEVICH2003279}, the asymptotic behavior of $K(s;x,x)$ goes as $s^{(k-m)/2}$ with $2$ corresponding to the degree of the operator, $m$ being the dimension of the manifold, and $k \in \mathbb{Z}^+$ being the $k$th leading term. For manifolds without imposed boundaries, it can be shown that only even $k$ contribute; relabeling coincides with the standard notation at~\cite{parker2009quantum} adopted in this paper. The coefficients read
\begin{align}
    E_{0}&=\mathbb{1}\,,\quad E_{1}=-\frac{\mathbb{1}}{12}R\,, \\
    E_{2}&= \frac{1}{12}\Lambda^{\mu \nu}\Lambda_{\mu \nu} +\frac{\mathbb{1}}{180}[R^{\mu \nu \rho \sigma}R_{\mu \nu \rho \sigma}-R^{\mu \nu}R_{\mu \nu}] \nonumber \\
    &\quad - \frac{1}{6}\nabla_{\mu}\nabla^{\mu}\left(\tfrac{1}{5}R -X\right)\mathbb{1} +\frac{1}{2}(\tfrac{1}{6}R-X)^2 \mathbb{1}\,,
\end{align}
with $\Lambda_{\mu \nu}= [\nabla_\mu ,\nabla_\nu] = -iR^{a b}_{\hphantom{ab} \mu \nu}\Sigma_{ab}$, where $\Sigma_{ab}=-\frac{i}{8}[\gamma_{a},\gamma_{b}]$ corresponds to the generators of the Lorentz group, and $X=\tfrac{1}{4}R$ for the elliptic Dirac operator.

\subsection{D. Estimation of the critical temperature}
\label{app:Estimation_Tc}

Obtaining a numerical value for $T_{\rm c}$ is equivalent to finding the zero temperature value of the effective surface tension, i.e., $\sigma$. The fermionic correction of the surface tension $\delta \sigma_{\rm eff}$ and elastic modulus $\kappa^{{}_{(2)}}_{G}$ are both negative and temperature dependent. Therefore attaining real values for the cylinder and sphere radii requires $\sigma$ to be bounded, $\delta \sigma_{\rm eff} (T_{\rm cyl})< \sigma < \delta \sigma_{\rm eff} (T_{\rm sph})$ for $T_{\rm sph} > T_{\rm cyl}$.
However, the bound is saturated when compared with observed values of the characteristic radii of the fullerene and nanotube structures at the temperatures required to produce them.
Imposing constraints on the radii of fullerene $r_{C60}<1.3\,$nm while satisfying $r_{\rm nt}<7\,$nm,
\begin{align}
    r_{C60} &= \left( \frac{2 \kappa^{{}_{(2)}}_{G}(T_{C60})}{\sigma + \delta \sigma_{\rm eff} (T_{C60})} \right)^{1/4} < 1.3\,\text{nm}, \nonumber \\
    r_{\rm nt} &= \sqrt{ \frac{\alpha/2}{\sigma + \delta \sigma_{\rm eff} (T_{\rm nt})} } < 7\,\text{nm} 
\end{align}
at their synthesis temperatures $T_{C60}=1000\,$K
and $T_{\rm nt}=720\,$K, respectively, within the range of known values in the literature~\cite{Sugai2000}. Hence, $\sigma_{\rm min} < \sigma < \sigma_{\rm max}$, where
\begin{align}
\sigma_{\rm min} &\coloneqq \frac{\alpha}{2 r_{\rm nt}^2} - \delta \sigma_{\rm eff} (T_{\rm nt})\;,\nonumber \\
\sigma_{\rm max} &\coloneqq 2\frac{\kappa^{{}_{(2)}}_G(T_{C60})}{r_{C60}^4} - \delta \sigma_{\rm eff} (T_{C60}).
\end{align}
This implies yet a stricter bound on the critical temperature $T_{\rm c}$,
\begin{equation}
    \left(\frac{\sigma_{\rm min} (\hbar v_{F})^{2}}{3g_{v}g_{s}\zeta(3){k_{B}^{3}}}\right)^{\frac{1}{3}} < T_{\rm c} < \left(\frac{\sigma_{\rm max}(\hbar v_{F})^{2}}{3g_{v}g_{s}\zeta(3){k_{B}^{3}}}\right)^{\frac{1}{3}},
\end{equation}
which translates to $955.201\,{\rm K} < T_{\rm c} < 960.312\,{\rm K}$; for Fig.\ref{fig:phase_diagram} we set the critical temperature to the average value $T_{\rm c} = 957.757\,$K. The numerical values for both $T_{\rm min}$ and $T_{c}$ are obtained using $\alpha=1.44~{\rm eV}$~\cite{YujieWei} and Fermi velocity $v_F = 0.85 \times 10^{6} \,{\rm m/s}$~\cite{hwang2012fermi}.

\bibliography{references}

\begin{thebibliography}{46}%
\makeatletter
\providecommand \@ifxundefined [1]{%
 \@ifx{#1\undefined}
}%
\providecommand \@ifnum [1]{%
 \ifnum #1\expandafter \@firstoftwo
 \else \expandafter \@secondoftwo
 \fi
}%
\providecommand \@ifx [1]{%
 \ifx #1\expandafter \@firstoftwo
 \else \expandafter \@secondoftwo
 \fi
}%
\providecommand \natexlab [1]{#1}%
\providecommand \enquote  [1]{``#1''}%
\providecommand \bibnamefont  [1]{#1}%
\providecommand \bibfnamefont [1]{#1}%
\providecommand \citenamefont [1]{#1}%
\providecommand \href@noop [0]{\@secondoftwo}%
\providecommand \href [0]{\begingroup \@sanitize@url \@href}%
\providecommand \@href[1]{\@@startlink{#1}\@@href}%
\providecommand \@@href[1]{\endgroup#1\@@endlink}%
\providecommand \@sanitize@url [0]{\catcode `\\12\catcode `\$12\catcode
  `\&12\catcode `\#12\catcode `\^12\catcode `\_12\catcode `\%12\relax}%
\providecommand \@@startlink[1]{}%
\providecommand \@@endlink[0]{}%
\providecommand \url  [0]{\begingroup\@sanitize@url \@url }%
\providecommand \@url [1]{\endgroup\@href {#1}{\urlprefix }}%
\providecommand \urlprefix  [0]{URL }%
\providecommand \Eprint [0]{\href }%
\providecommand \doibase [0]{https://doi.org/}%
\providecommand \selectlanguage [0]{\@gobble}%
\providecommand \bibinfo  [0]{\@secondoftwo}%
\providecommand \bibfield  [0]{\@secondoftwo}%
\providecommand \translation [1]{[#1]}%
\providecommand \BibitemOpen [0]{}%
\providecommand \bibitemStop [0]{}%
\providecommand \bibitemNoStop [0]{.\EOS\space}%
\providecommand \EOS [0]{\spacefactor3000\relax}%
\providecommand \BibitemShut  [1]{\csname bibitem#1\endcsname}%
\let\auto@bib@innerbib\@empty
\bibitem [{\citenamefont {Castro~Neto}\ \emph {et~al.}(2009)\citenamefont
  {Castro~Neto}, \citenamefont {Guinea}, \citenamefont {Peres}, \citenamefont
  {Novoselov},\ and\ \citenamefont {Geim}}]{RevModPhys.81.109}%
  \BibitemOpen
  \bibfield  {author} {\bibinfo {author} {\bibfnamefont {A.~H.}\ \bibnamefont
  {Castro~Neto}}, \bibinfo {author} {\bibfnamefont {F.}~\bibnamefont {Guinea}},
  \bibinfo {author} {\bibfnamefont {N.~M.~R.}\ \bibnamefont {Peres}}, \bibinfo
  {author} {\bibfnamefont {K.~S.}\ \bibnamefont {Novoselov}},\ and\ \bibinfo
  {author} {\bibfnamefont {A.~K.}\ \bibnamefont {Geim}},\ }\href
  {https://doi.org/10.1103/RevModPhys.81.109} {\bibfield  {journal} {\bibinfo
  {journal} {Rev. Mod. Phys.}\ }\textbf {\bibinfo {volume} {81}},\ \bibinfo
  {pages} {109} (\bibinfo {year} {2009})}\BibitemShut {NoStop}%
\bibitem [{\citenamefont {Mélinon}(2021)}]{nano11071694}%
  \BibitemOpen
  \bibfield  {author} {\bibinfo {author} {\bibfnamefont {P.}~\bibnamefont
  {Mélinon}},\ }\bibfield  {journal} {\bibinfo  {journal} {Nanomaterials}\
  }\textbf {\bibinfo {volume} {11}},\ \href
  {https://doi.org/10.3390/nano11071694} {10.3390/nano11071694} (\bibinfo
  {year} {2021})\BibitemShut {NoStop}%
\bibitem [{\citenamefont {Novoselov}\ \emph {et~al.}(2005)\citenamefont
  {Novoselov}, \citenamefont {Geim}, \citenamefont {Morozov}, \citenamefont
  {Jiang}, \citenamefont {Katsnelson}, \citenamefont {Grigorieva},
  \citenamefont {Dubonos},\ and\ \citenamefont {Firsov}}]{Novoselov2}%
  \BibitemOpen
  \bibfield  {author} {\bibinfo {author} {\bibfnamefont {K.~S.}\ \bibnamefont
  {Novoselov}}, \bibinfo {author} {\bibfnamefont {A.~K.}\ \bibnamefont {Geim}},
  \bibinfo {author} {\bibfnamefont {S.~V.}\ \bibnamefont {Morozov}}, \bibinfo
  {author} {\bibfnamefont {D.}~\bibnamefont {Jiang}}, \bibinfo {author}
  {\bibfnamefont {M.~I.}\ \bibnamefont {Katsnelson}}, \bibinfo {author}
  {\bibfnamefont {I.~V.}\ \bibnamefont {Grigorieva}}, \bibinfo {author}
  {\bibfnamefont {S.~V.}\ \bibnamefont {Dubonos}},\ and\ \bibinfo {author}
  {\bibfnamefont {A.~A.}\ \bibnamefont {Firsov}},\ }\href
  {https://doi.org/10.1038/nature04233} {\bibfield  {journal} {\bibinfo
  {journal} {Nature}\ }\textbf {\bibinfo {volume} {438}},\ \bibinfo {pages}
  {197} (\bibinfo {year} {2005})}\BibitemShut {NoStop}%
\bibitem [{\citenamefont {Naumis}\ \emph {et~al.}(2017)\citenamefont {Naumis},
  \citenamefont {Barraza-Lopez}, \citenamefont {Oliva-Leyva},\ and\
  \citenamefont {Terrones}}]{Naumis_2017}%
  \BibitemOpen
  \bibfield  {author} {\bibinfo {author} {\bibfnamefont {G.~G.}\ \bibnamefont
  {Naumis}}, \bibinfo {author} {\bibfnamefont {S.}~\bibnamefont
  {Barraza-Lopez}}, \bibinfo {author} {\bibfnamefont {M.}~\bibnamefont
  {Oliva-Leyva}},\ and\ \bibinfo {author} {\bibfnamefont {H.}~\bibnamefont
  {Terrones}},\ }\href {https://doi.org/10.1088/1361-6633/aa74ef} {\bibfield
  {journal} {\bibinfo  {journal} {Reports on Progress in Physics}\ }\textbf
  {\bibinfo {volume} {80}},\ \bibinfo {pages} {096501} (\bibinfo {year}
  {2017})}\BibitemShut {NoStop}%
\bibitem [{\citenamefont {Naumis}\ \emph {et~al.}(2023)\citenamefont {Naumis},
  \citenamefont {Herrera}, \citenamefont {Poudel}, \citenamefont {Nakamura},\
  and\ \citenamefont {Barraza-Lopez}}]{Naumis_2024}%
  \BibitemOpen
  \bibfield  {author} {\bibinfo {author} {\bibfnamefont {G.~G.}\ \bibnamefont
  {Naumis}}, \bibinfo {author} {\bibfnamefont {S.~A.}\ \bibnamefont {Herrera}},
  \bibinfo {author} {\bibfnamefont {S.~P.}\ \bibnamefont {Poudel}}, \bibinfo
  {author} {\bibfnamefont {H.}~\bibnamefont {Nakamura}},\ and\ \bibinfo
  {author} {\bibfnamefont {S.}~\bibnamefont {Barraza-Lopez}},\ }\href
  {https://doi.org/10.1088/1361-6633/ad06db} {\bibfield  {journal} {\bibinfo
  {journal} {Reports on Progress in Physics}\ }\textbf {\bibinfo {volume}
  {87}},\ \bibinfo {pages} {016502} (\bibinfo {year} {2023})}\BibitemShut
  {NoStop}%
\bibitem [{\citenamefont {Levy}\ \emph {et~al.}(2010)\citenamefont {Levy},
  \citenamefont {Burke}, \citenamefont {Meaker}, \citenamefont {Panlasigui},
  \citenamefont {Zettl}, \citenamefont {Guinea}, \citenamefont {Neto},\ and\
  \citenamefont {Crommie}}]{doi:10.1126/science.1191700}%
  \BibitemOpen
  \bibfield  {author} {\bibinfo {author} {\bibfnamefont {N.}~\bibnamefont
  {Levy}}, \bibinfo {author} {\bibfnamefont {S.~A.}\ \bibnamefont {Burke}},
  \bibinfo {author} {\bibfnamefont {K.~L.}\ \bibnamefont {Meaker}}, \bibinfo
  {author} {\bibfnamefont {M.}~\bibnamefont {Panlasigui}}, \bibinfo {author}
  {\bibfnamefont {A.}~\bibnamefont {Zettl}}, \bibinfo {author} {\bibfnamefont
  {F.}~\bibnamefont {Guinea}}, \bibinfo {author} {\bibfnamefont {A.~H.~C.}\
  \bibnamefont {Neto}},\ and\ \bibinfo {author} {\bibfnamefont {M.~F.}\
  \bibnamefont {Crommie}},\ }\href {https://doi.org/10.1126/science.1191700}
  {\bibfield  {journal} {\bibinfo  {journal} {Science}\ }\textbf {\bibinfo
  {volume} {329}},\ \bibinfo {pages} {544} (\bibinfo {year} {2010})},\ \Eprint
  {https://arxiv.org/abs/https://www.science.org/doi/pdf/10.1126/science.1191700}
  {https://www.science.org/doi/pdf/10.1126/science.1191700} \BibitemShut
  {NoStop}%
\bibitem [{\citenamefont {Castro-Villarreal}\ and\ \citenamefont
  {Ruiz-S\'anchez}(2017)}]{Castro2017}%
  \BibitemOpen
  \bibfield  {author} {\bibinfo {author} {\bibfnamefont {P.}~\bibnamefont
  {Castro-Villarreal}}\ and\ \bibinfo {author} {\bibfnamefont {R.}~\bibnamefont
  {Ruiz-S\'anchez}},\ }\href {https://doi.org/10.1103/PhysRevB.95.125432}
  {\bibfield  {journal} {\bibinfo  {journal} {Phys. Rev. B}\ }\textbf {\bibinfo
  {volume} {95}},\ \bibinfo {pages} {125432} (\bibinfo {year}
  {2017})}\BibitemShut {NoStop}%
\bibitem [{\citenamefont {Morales}\ and\ \citenamefont
  {Copinger}(2023)}]{Morales2023}%
  \BibitemOpen
  \bibfield  {author} {\bibinfo {author} {\bibfnamefont {P.~A.}\ \bibnamefont
  {Morales}}\ and\ \bibinfo {author} {\bibfnamefont {P.}~\bibnamefont
  {Copinger}},\ }\href {https://doi.org/10.1103/PhysRevB.107.075432} {\bibfield
   {journal} {\bibinfo  {journal} {Phys. Rev. B}\ }\textbf {\bibinfo {volume}
  {107}},\ \bibinfo {pages} {075432} (\bibinfo {year} {2023})}\BibitemShut
  {NoStop}%
\bibitem [{\citenamefont {Kun}\ \emph {et~al.}(2019)\citenamefont {Kun},
  \citenamefont {Kukucska}, \citenamefont {Dobrik}, \citenamefont {Koltai},
  \citenamefont {K{\"u}rti}, \citenamefont {Bir{\'o}}, \citenamefont
  {Tapaszt{\'o}},\ and\ \citenamefont {Nemes-Incze}}]{Kun2019}%
  \BibitemOpen
  \bibfield  {author} {\bibinfo {author} {\bibfnamefont {P.}~\bibnamefont
  {Kun}}, \bibinfo {author} {\bibfnamefont {G.}~\bibnamefont {Kukucska}},
  \bibinfo {author} {\bibfnamefont {G.}~\bibnamefont {Dobrik}}, \bibinfo
  {author} {\bibfnamefont {J.}~\bibnamefont {Koltai}}, \bibinfo {author}
  {\bibfnamefont {J.}~\bibnamefont {K{\"u}rti}}, \bibinfo {author}
  {\bibfnamefont {L.~P.}\ \bibnamefont {Bir{\'o}}}, \bibinfo {author}
  {\bibfnamefont {L.}~\bibnamefont {Tapaszt{\'o}}},\ and\ \bibinfo {author}
  {\bibfnamefont {P.}~\bibnamefont {Nemes-Incze}},\ }\href
  {https://doi.org/10.1038/s41699-019-0094-6} {\bibfield  {journal} {\bibinfo
  {journal} {npj 2D Materials and Applications}\ }\textbf {\bibinfo {volume}
  {3}},\ \bibinfo {pages} {11} (\bibinfo {year} {2019})}\BibitemShut {NoStop}%
\bibitem [{\citenamefont {Roberts}\ and\ \citenamefont
  {Wiseman}(2022{\natexlab{a}})}]{MattWiseman1}%
  \BibitemOpen
  \bibfield  {author} {\bibinfo {author} {\bibfnamefont {M.~M.}\ \bibnamefont
  {Roberts}}\ and\ \bibinfo {author} {\bibfnamefont {T.}~\bibnamefont
  {Wiseman}},\ }\href {https://doi.org/10.1103/PhysRevB.105.195412} {\bibfield
  {journal} {\bibinfo  {journal} {Phys. Rev. B}\ }\textbf {\bibinfo {volume}
  {105}},\ \bibinfo {pages} {195412} (\bibinfo {year}
  {2022}{\natexlab{a}})}\BibitemShut {NoStop}%
\bibitem [{\citenamefont {Roberts}\ and\ \citenamefont
  {Wiseman}(2024)}]{MattWiseman2}%
  \BibitemOpen
  \bibfield  {author} {\bibinfo {author} {\bibfnamefont {M.~M.}\ \bibnamefont
  {Roberts}}\ and\ \bibinfo {author} {\bibfnamefont {T.}~\bibnamefont
  {Wiseman}},\ }\href {https://doi.org/10.1103/PhysRevB.109.045425} {\bibfield
  {journal} {\bibinfo  {journal} {Phys. Rev. B}\ }\textbf {\bibinfo {volume}
  {109}},\ \bibinfo {pages} {045425} (\bibinfo {year} {2024})}\BibitemShut
  {NoStop}%
\bibitem [{\citenamefont {Iorio}\ and\ \citenamefont
  {Pais}(2022)}]{CommentPaper}%
  \BibitemOpen
  \bibfield  {author} {\bibinfo {author} {\bibfnamefont {A.}~\bibnamefont
  {Iorio}}\ and\ \bibinfo {author} {\bibfnamefont {P.}~\bibnamefont {Pais}},\
  }\href {https://doi.org/10.1103/PhysRevB.106.157401} {\bibfield  {journal}
  {\bibinfo  {journal} {Phys. Rev. B}\ }\textbf {\bibinfo {volume} {106}},\
  \bibinfo {pages} {157401} (\bibinfo {year} {2022})}\BibitemShut {NoStop}%
\bibitem [{\citenamefont {Roberts}\ and\ \citenamefont
  {Wiseman}(2022{\natexlab{b}})}]{ReplyToComment}%
  \BibitemOpen
  \bibfield  {author} {\bibinfo {author} {\bibfnamefont {M.~M.}\ \bibnamefont
  {Roberts}}\ and\ \bibinfo {author} {\bibfnamefont {T.}~\bibnamefont
  {Wiseman}},\ }\href {https://doi.org/10.1103/PhysRevB.106.157402} {\bibfield
  {journal} {\bibinfo  {journal} {Phys. Rev. B}\ }\textbf {\bibinfo {volume}
  {106}},\ \bibinfo {pages} {157402} (\bibinfo {year}
  {2022}{\natexlab{b}})}\BibitemShut {NoStop}%
\bibitem [{\citenamefont {Espinosa-Champo}\ \emph {et~al.}(2024)\citenamefont
  {Espinosa-Champo}, \citenamefont {Naumis},\ and\ \citenamefont
  {Castro-Villarreal}}]{Castro2024}%
  \BibitemOpen
  \bibfield  {author} {\bibinfo {author} {\bibfnamefont {A.~d.~J.}\
  \bibnamefont {Espinosa-Champo}}, \bibinfo {author} {\bibfnamefont {G.~G.}\
  \bibnamefont {Naumis}},\ and\ \bibinfo {author} {\bibfnamefont
  {P.}~\bibnamefont {Castro-Villarreal}},\ }\href
  {https://doi.org/10.1103/PhysRevB.110.035421} {\bibfield  {journal} {\bibinfo
   {journal} {Phys. Rev. B}\ }\textbf {\bibinfo {volume} {110}},\ \bibinfo
  {pages} {035421} (\bibinfo {year} {2024})}\BibitemShut {NoStop}%
\bibitem [{\citenamefont {Srisuma}\ \emph {et~al.}(2021)\citenamefont
  {Srisuma}, \citenamefont {Suwattanapongtada}, \citenamefont {Opasanon},
  \citenamefont {Charoensuppanimit}, \citenamefont {Kerdnawee}, \citenamefont
  {Termvidchakorn}, \citenamefont {Tanthapanichakoon},\ and\ \citenamefont
  {Charinpanitku}}]{SRISUMA20211373}%
  \BibitemOpen
  \bibfield  {author} {\bibinfo {author} {\bibfnamefont {P.}~\bibnamefont
  {Srisuma}}, \bibinfo {author} {\bibfnamefont {N.}~\bibnamefont
  {Suwattanapongtada}}, \bibinfo {author} {\bibfnamefont {N.}~\bibnamefont
  {Opasanon}}, \bibinfo {author} {\bibfnamefont {P.}~\bibnamefont
  {Charoensuppanimit}}, \bibinfo {author} {\bibfnamefont {K.}~\bibnamefont
  {Kerdnawee}}, \bibinfo {author} {\bibfnamefont {C.}~\bibnamefont
  {Termvidchakorn}}, \bibinfo {author} {\bibfnamefont {W.}~\bibnamefont
  {Tanthapanichakoon}},\ and\ \bibinfo {author} {\bibfnamefont
  {T.}~\bibnamefont {Charinpanitku}},\ }\href
  {https://doi.org/https://doi.org/10.1016/j.jestch.2021.03.006} {\bibfield
  {journal} {\bibinfo  {journal} {Engineering Science and Technology, an
  International Journal}\ }\textbf {\bibinfo {volume} {24}},\ \bibinfo {pages}
  {1373} (\bibinfo {year} {2021})}\BibitemShut {NoStop}%
\bibitem [{\citenamefont {Chuvilin}\ \emph {et~al.}(2010)\citenamefont
  {Chuvilin}, \citenamefont {Kaiser}, \citenamefont {Bichoutskaia},
  \citenamefont {Besley},\ and\ \citenamefont {Khlobystov}}]{Chuvilin2010}%
  \BibitemOpen
  \bibfield  {author} {\bibinfo {author} {\bibfnamefont {A.}~\bibnamefont
  {Chuvilin}}, \bibinfo {author} {\bibfnamefont {U.}~\bibnamefont {Kaiser}},
  \bibinfo {author} {\bibfnamefont {E.}~\bibnamefont {Bichoutskaia}}, \bibinfo
  {author} {\bibfnamefont {N.~A.}\ \bibnamefont {Besley}},\ and\ \bibinfo
  {author} {\bibfnamefont {A.~N.}\ \bibnamefont {Khlobystov}},\ }\href
  {https://doi.org/10.1038/nchem.644} {\bibfield  {journal} {\bibinfo
  {journal} {Nature Chemistry}\ }\textbf {\bibinfo {volume} {2}},\ \bibinfo
  {pages} {450} (\bibinfo {year} {2010})}\BibitemShut {NoStop}%
\bibitem [{\citenamefont {Fischetti}\ \emph {et~al.}(2018)\citenamefont
  {Fischetti}, \citenamefont {Wallis},\ and\ \citenamefont
  {Wiseman}}]{PhysRevLett.120.261601}%
  \BibitemOpen
  \bibfield  {author} {\bibinfo {author} {\bibfnamefont {S.}~\bibnamefont
  {Fischetti}}, \bibinfo {author} {\bibfnamefont {L.}~\bibnamefont {Wallis}},\
  and\ \bibinfo {author} {\bibfnamefont {T.}~\bibnamefont {Wiseman}},\ }\href
  {https://doi.org/10.1103/PhysRevLett.120.261601} {\bibfield  {journal}
  {\bibinfo  {journal} {Phys. Rev. Lett.}\ }\textbf {\bibinfo {volume} {120}},\
  \bibinfo {pages} {261601} (\bibinfo {year} {2018})}\BibitemShut {NoStop}%
\bibitem [{\citenamefont {Kolesnikov}\ and\ \citenamefont
  {Osipov}(2006)}]{kolesnikov2006continuum}%
  \BibitemOpen
  \bibfield  {author} {\bibinfo {author} {\bibfnamefont {D.}~\bibnamefont
  {Kolesnikov}}\ and\ \bibinfo {author} {\bibfnamefont {V.}~\bibnamefont
  {Osipov}},\ }\href@noop {} {\bibfield  {journal} {\bibinfo  {journal} {The
  European Physical Journal B-Condensed Matter and Complex Systems}\ }\textbf
  {\bibinfo {volume} {49}},\ \bibinfo {pages} {465} (\bibinfo {year}
  {2006})}\BibitemShut {NoStop}%
\bibitem [{\citenamefont {Kroto}\ \emph {et~al.}(1985)\citenamefont {Kroto},
  \citenamefont {Heath}, \citenamefont {O’Brien}, \citenamefont {Curl},\ and\
  \citenamefont {Smalley}}]{Kroto}%
  \BibitemOpen
  \bibfield  {author} {\bibinfo {author} {\bibfnamefont {H.~W.}\ \bibnamefont
  {Kroto}}, \bibinfo {author} {\bibfnamefont {J.~R.}\ \bibnamefont {Heath}},
  \bibinfo {author} {\bibfnamefont {S.~C.}\ \bibnamefont {O’Brien}}, \bibinfo
  {author} {\bibfnamefont {R.~F.}\ \bibnamefont {Curl}},\ and\ \bibinfo
  {author} {\bibfnamefont {R.~E.}\ \bibnamefont {Smalley}},\ }\href
  {https://doi.org/10.1038/318162a0} {\bibfield  {journal} {\bibinfo  {journal}
  {Nature}\ }\textbf {\bibinfo {volume} {318}},\ \bibinfo {pages} {162}
  (\bibinfo {year} {1985})}\BibitemShut {NoStop}%
\bibitem [{\citenamefont {Iijima}(1991)}]{Lijima}%
  \BibitemOpen
  \bibfield  {author} {\bibinfo {author} {\bibfnamefont {S.}~\bibnamefont
  {Iijima}},\ }\href {https://doi.org/10.1038/354056a0} {\bibfield  {journal}
  {\bibinfo  {journal} {Nature}\ }\textbf {\bibinfo {volume} {354}},\ \bibinfo
  {pages} {56} (\bibinfo {year} {1991})}\BibitemShut {NoStop}%
\bibitem [{\citenamefont {Terrones}\ and\ \citenamefont
  {Mackay}(1992)}]{Terrones}%
  \BibitemOpen
  \bibfield  {author} {\bibinfo {author} {\bibfnamefont {H.}~\bibnamefont
  {Terrones}}\ and\ \bibinfo {author} {\bibfnamefont {A.}~\bibnamefont
  {Mackay}},\ }\href
  {https://doi.org/https://doi.org/10.1016/0008-6223(92)90066-6} {\bibfield
  {journal} {\bibinfo  {journal} {Carbon}\ }\textbf {\bibinfo {volume} {30}},\
  \bibinfo {pages} {1251} (\bibinfo {year} {1992})}\BibitemShut {NoStop}%
\bibitem [{\citenamefont {Terrones}\ \emph {et~al.}(2010)\citenamefont
  {Terrones}, \citenamefont {Botello-Méndez}, \citenamefont {Campos-Delgado},
  \citenamefont {López-Urías}, \citenamefont {Vega-Cantú}, \citenamefont
  {Rodríguez-Macías}, \citenamefont {Elías}, \citenamefont
  {Muñoz-Sandoval}, \citenamefont {Cano-Márquez}, \citenamefont {Charlier},\
  and\ \citenamefont {Terrones}}]{TERRONES2010351}%
  \BibitemOpen
  \bibfield  {author} {\bibinfo {author} {\bibfnamefont {M.}~\bibnamefont
  {Terrones}}, \bibinfo {author} {\bibfnamefont {A.~R.}\ \bibnamefont
  {Botello-Méndez}}, \bibinfo {author} {\bibfnamefont {J.}~\bibnamefont
  {Campos-Delgado}}, \bibinfo {author} {\bibfnamefont {F.}~\bibnamefont
  {López-Urías}}, \bibinfo {author} {\bibfnamefont {Y.~I.}\ \bibnamefont
  {Vega-Cantú}}, \bibinfo {author} {\bibfnamefont {F.~J.}\ \bibnamefont
  {Rodríguez-Macías}}, \bibinfo {author} {\bibfnamefont {A.~L.}\ \bibnamefont
  {Elías}}, \bibinfo {author} {\bibfnamefont {E.}~\bibnamefont
  {Muñoz-Sandoval}}, \bibinfo {author} {\bibfnamefont {A.~G.}\ \bibnamefont
  {Cano-Márquez}}, \bibinfo {author} {\bibfnamefont {J.-C.}\ \bibnamefont
  {Charlier}},\ and\ \bibinfo {author} {\bibfnamefont {H.}~\bibnamefont
  {Terrones}},\ }\href
  {https://doi.org/https://doi.org/10.1016/j.nantod.2010.06.010} {\bibfield
  {journal} {\bibinfo  {journal} {Nano Today}\ }\textbf {\bibinfo {volume}
  {5}},\ \bibinfo {pages} {351} (\bibinfo {year} {2010})}\BibitemShut {NoStop}%
\bibitem [{\citenamefont {Canham}(1970)}]{CANHAM197061}%
  \BibitemOpen
  \bibfield  {author} {\bibinfo {author} {\bibfnamefont {P.}~\bibnamefont
  {Canham}},\ }\href
  {https://doi.org/https://doi.org/10.1016/S0022-5193(70)80032-7} {\bibfield
  {journal} {\bibinfo  {journal} {Journal of Theoretical Biology}\ }\textbf
  {\bibinfo {volume} {26}},\ \bibinfo {pages} {61} (\bibinfo {year}
  {1970})}\BibitemShut {NoStop}%
\bibitem [{\citenamefont {Helfrich}(1973)}]{Helfrich1973}%
  \BibitemOpen
  \bibfield  {author} {\bibinfo {author} {\bibfnamefont {W.}~\bibnamefont
  {Helfrich}},\ }\href@noop {} {\bibfield  {journal} {\bibinfo  {journal} {Z
  Naturforsch C}\ }\textbf {\bibinfo {volume} {28}},\ \bibinfo {pages} {693}
  (\bibinfo {year} {1973})}\BibitemShut {NoStop}%
\bibitem [{\citenamefont {Kim}\ and\ \citenamefont {Neto}(2008)}]{Kim_2008}%
  \BibitemOpen
  \bibfield  {author} {\bibinfo {author} {\bibfnamefont {E.-A.}\ \bibnamefont
  {Kim}}\ and\ \bibinfo {author} {\bibfnamefont {A.~H.~C.}\ \bibnamefont
  {Neto}},\ }\href {https://doi.org/10.1209/0295-5075/84/57007} {\bibfield
  {journal} {\bibinfo  {journal} {Europhysics Letters}\ }\textbf {\bibinfo
  {volume} {84}},\ \bibinfo {pages} {57007} (\bibinfo {year}
  {2008})}\BibitemShut {NoStop}%
\bibitem [{\citenamefont {Deserno}(2015)}]{DESERNO201511}%
  \BibitemOpen
  \bibfield  {author} {\bibinfo {author} {\bibfnamefont {M.}~\bibnamefont
  {Deserno}},\ }\href
  {https://doi.org/https://doi.org/10.1016/j.chemphyslip.2014.05.001}
  {\bibfield  {journal} {\bibinfo  {journal} {Chemistry and Physics of Lipids}\
  }\textbf {\bibinfo {volume} {185}},\ \bibinfo {pages} {11} (\bibinfo {year}
  {2015})},\ \bibinfo {note} {membrane mechanochemistry: From the molecular to
  the cellular scale}\BibitemShut {NoStop}%
\bibitem [{\citenamefont {do~Carmo}(1976)}]{DoCarmo}%
  \BibitemOpen
  \bibfield  {author} {\bibinfo {author} {\bibfnamefont {M.~P.}\ \bibnamefont
  {do~Carmo}},\ }\href@noop {} {\emph {\bibinfo {title} {Differential geometry
  of curves and surfaces.}}}\ (\bibinfo  {publisher} {Prentice Hall},\ \bibinfo
  {year} {1976})\ pp.\ \bibinfo {pages} {I--VIII, 1--503}\BibitemShut {NoStop}%
\bibitem [{\citenamefont {Wei}\ \emph {et~al.}(2013)\citenamefont {Wei},
  \citenamefont {Wang}, \citenamefont {Wu}, \citenamefont {Yang},\ and\
  \citenamefont {Dunn}}]{YujieWei}%
  \BibitemOpen
  \bibfield  {author} {\bibinfo {author} {\bibfnamefont {Y.}~\bibnamefont
  {Wei}}, \bibinfo {author} {\bibfnamefont {B.}~\bibnamefont {Wang}}, \bibinfo
  {author} {\bibfnamefont {J.}~\bibnamefont {Wu}}, \bibinfo {author}
  {\bibfnamefont {R.}~\bibnamefont {Yang}},\ and\ \bibinfo {author}
  {\bibfnamefont {M.~L.}\ \bibnamefont {Dunn}},\ }\href
  {https://doi.org/10.1021/nl303168w} {\bibfield  {journal} {\bibinfo
  {journal} {Nano Letters}\ }\textbf {\bibinfo {volume} {13}},\ \bibinfo
  {pages} {26} (\bibinfo {year} {2013})},\ \bibinfo {note} {pMID: 23214980},\
  \Eprint {https://arxiv.org/abs/https://doi.org/10.1021/nl303168w}
  {https://doi.org/10.1021/nl303168w} \BibitemShut {NoStop}%
\bibitem [{\citenamefont {Ashino}\ \emph
  {et~al.}(2021{\natexlab{a}})\citenamefont {Ashino}, \citenamefont {Nishioka},
  \citenamefont {Hayashi},\ and\ \citenamefont {Wiesendanger}}]{Ashino1}%
  \BibitemOpen
  \bibfield  {author} {\bibinfo {author} {\bibfnamefont {M.}~\bibnamefont
  {Ashino}}, \bibinfo {author} {\bibfnamefont {K.}~\bibnamefont {Nishioka}},
  \bibinfo {author} {\bibfnamefont {K.}~\bibnamefont {Hayashi}},\ and\ \bibinfo
  {author} {\bibfnamefont {R.}~\bibnamefont {Wiesendanger}},\ }\href
  {https://doi.org/10.1103/PhysRevLett.126.146101} {\bibfield  {journal}
  {\bibinfo  {journal} {Phys. Rev. Lett.}\ }\textbf {\bibinfo {volume} {126}},\
  \bibinfo {pages} {146101} (\bibinfo {year} {2021}{\natexlab{a}})}\BibitemShut
  {NoStop}%
\bibitem [{\citenamefont {Ashino}\ \emph
  {et~al.}(2021{\natexlab{b}})\citenamefont {Ashino}, \citenamefont {Nishioka},
  \citenamefont {Hayashi},\ and\ \citenamefont {Wiesendanger}}]{Ashino2}%
  \BibitemOpen
  \bibfield  {author} {\bibinfo {author} {\bibfnamefont {M.}~\bibnamefont
  {Ashino}}, \bibinfo {author} {\bibfnamefont {K.}~\bibnamefont {Nishioka}},
  \bibinfo {author} {\bibfnamefont {K.}~\bibnamefont {Hayashi}},\ and\ \bibinfo
  {author} {\bibfnamefont {R.}~\bibnamefont {Wiesendanger}},\ }\href
  {https://doi.org/10.1103/PhysRevB.104.085407} {\bibfield  {journal} {\bibinfo
   {journal} {Phys. Rev. B}\ }\textbf {\bibinfo {volume} {104}},\ \bibinfo
  {pages} {085407} (\bibinfo {year} {2021}{\natexlab{b}})}\BibitemShut
  {NoStop}%
\bibitem [{\citenamefont {Iorio}\ and\ \citenamefont
  {Lambiase}(2012)}]{IORIO2012}%
  \BibitemOpen
  \bibfield  {author} {\bibinfo {author} {\bibfnamefont {A.}~\bibnamefont
  {Iorio}}\ and\ \bibinfo {author} {\bibfnamefont {G.}~\bibnamefont
  {Lambiase}},\ }\href
  {https://doi.org/https://doi.org/10.1016/j.physletb.2012.08.023} {\bibfield
  {journal} {\bibinfo  {journal} {Physics Letters B}\ }\textbf {\bibinfo
  {volume} {716}},\ \bibinfo {pages} {334} (\bibinfo {year}
  {2012})}\BibitemShut {NoStop}%
\bibitem [{\citenamefont {Iorio}\ and\ \citenamefont
  {Lambiase}(2014)}]{Iorio2014}%
  \BibitemOpen
  \bibfield  {author} {\bibinfo {author} {\bibfnamefont {A.}~\bibnamefont
  {Iorio}}\ and\ \bibinfo {author} {\bibfnamefont {G.}~\bibnamefont
  {Lambiase}},\ }\href {https://doi.org/10.1103/PhysRevD.90.025006} {\bibfield
  {journal} {\bibinfo  {journal} {Phys. Rev. D}\ }\textbf {\bibinfo {volume}
  {90}},\ \bibinfo {pages} {025006} (\bibinfo {year} {2014})}\BibitemShut
  {NoStop}%
\bibitem [{\citenamefont {Altland}\ and\ \citenamefont
  {Simons}(2010)}]{Altland}%
  \BibitemOpen
  \bibfield  {author} {\bibinfo {author} {\bibfnamefont {A.}~\bibnamefont
  {Altland}}\ and\ \bibinfo {author} {\bibfnamefont {B.~D.}\ \bibnamefont
  {Simons}},\ }\href@noop {} {\emph {\bibinfo {title} {Condensed matter field
  theory}}}\ (\bibinfo  {publisher} {Cambridge University Press},\ \bibinfo
  {year} {2010})\BibitemShut {NoStop}%
\bibitem [{\citenamefont {Parker}\ and\ \citenamefont
  {Toms}(2009)}]{parker2009quantum}%
  \BibitemOpen
  \bibfield  {author} {\bibinfo {author} {\bibfnamefont {L.}~\bibnamefont
  {Parker}}\ and\ \bibinfo {author} {\bibfnamefont {D.}~\bibnamefont {Toms}},\
  }\href {https://books.google.co.jp/books?id=5nNuGMBBTjMC} {\emph {\bibinfo
  {title} {Quantum Field Theory in Curved Spacetime: Quantized Fields and
  Gravity}}},\ Cambridge Monographs on Mathematical Physics\ (\bibinfo
  {publisher} {Cambridge University Press},\ \bibinfo {year}
  {2009})\BibitemShut {NoStop}%
\bibitem [{\citenamefont {Vassilevich}(2003)}]{VASSILEVICH2003279}%
  \BibitemOpen
  \bibfield  {author} {\bibinfo {author} {\bibfnamefont {D.}~\bibnamefont
  {Vassilevich}},\ }\href
  {https://doi.org/https://doi.org/10.1016/j.physrep.2003.09.002} {\bibfield
  {journal} {\bibinfo  {journal} {Physics Reports}\ }\textbf {\bibinfo {volume}
  {388}},\ \bibinfo {pages} {279} (\bibinfo {year} {2003})}\BibitemShut
  {NoStop}%
\bibitem [{\citenamefont {Boschi-Filho}\ and\ \citenamefont
  {Natividade}(1992)}]{PhysRevD.46.5458}%
  \BibitemOpen
  \bibfield  {author} {\bibinfo {author} {\bibfnamefont {H.}~\bibnamefont
  {Boschi-Filho}}\ and\ \bibinfo {author} {\bibfnamefont {C.~P.}\ \bibnamefont
  {Natividade}},\ }\href {https://doi.org/10.1103/PhysRevD.46.5458} {\bibfield
  {journal} {\bibinfo  {journal} {Phys. Rev. D}\ }\textbf {\bibinfo {volume}
  {46}},\ \bibinfo {pages} {5458} (\bibinfo {year} {1992})}\BibitemShut
  {NoStop}%
\bibitem [{\citenamefont {Fischetti}\ \emph {et~al.}(2020)\citenamefont
  {Fischetti}, \citenamefont {Wallis},\ and\ \citenamefont
  {Wiseman}}]{fischetti2020does}%
  \BibitemOpen
  \bibfield  {author} {\bibinfo {author} {\bibfnamefont {S.}~\bibnamefont
  {Fischetti}}, \bibinfo {author} {\bibfnamefont {L.}~\bibnamefont {Wallis}},\
  and\ \bibinfo {author} {\bibfnamefont {T.}~\bibnamefont {Wiseman}},\
  }\href@noop {} {\bibfield  {journal} {\bibinfo  {journal} {Journal of High
  Energy Physics}\ }\textbf {\bibinfo {volume} {2020}},\ \bibinfo {pages} {1}
  (\bibinfo {year} {2020})}\BibitemShut {NoStop}%
\bibitem [{\citenamefont {Guven}(2004)}]{JemalGuven_2004}%
  \BibitemOpen
  \bibfield  {author} {\bibinfo {author} {\bibfnamefont {J.}~\bibnamefont
  {Guven}},\ }\href {https://doi.org/10.1088/0305-4470/37/28/L02} {\bibfield
  {journal} {\bibinfo  {journal} {Journal of Physics A: Mathematical and
  General}\ }\textbf {\bibinfo {volume} {37}},\ \bibinfo {pages} {L313}
  (\bibinfo {year} {2004})}\BibitemShut {NoStop}%
\bibitem [{\citenamefont {Capovilla}\ \emph {et~al.}(2003)\citenamefont
  {Capovilla}, \citenamefont {Guven},\ and\ \citenamefont
  {Santiago}}]{capovilla2003deformations}%
  \BibitemOpen
  \bibfield  {author} {\bibinfo {author} {\bibfnamefont {R.}~\bibnamefont
  {Capovilla}}, \bibinfo {author} {\bibfnamefont {J.}~\bibnamefont {Guven}},\
  and\ \bibinfo {author} {\bibfnamefont {J.}~\bibnamefont {Santiago}},\
  }\href@noop {} {\bibfield  {journal} {\bibinfo  {journal} {Journal of Physics
  A: Mathematical and General}\ }\textbf {\bibinfo {volume} {36}},\ \bibinfo
  {pages} {6281} (\bibinfo {year} {2003})}\BibitemShut {NoStop}%
\bibitem [{\citenamefont {Sugai}\ \emph {et~al.}(2000)\citenamefont {Sugai},
  \citenamefont {Omote}, \citenamefont {Bandow}, \citenamefont {Tanaka},\ and\
  \citenamefont {Shinohara}}]{Sugai2000}%
  \BibitemOpen
  \bibfield  {author} {\bibinfo {author} {\bibfnamefont {T.}~\bibnamefont
  {Sugai}}, \bibinfo {author} {\bibfnamefont {H.}~\bibnamefont {Omote}},
  \bibinfo {author} {\bibfnamefont {S.}~\bibnamefont {Bandow}}, \bibinfo
  {author} {\bibfnamefont {N.}~\bibnamefont {Tanaka}},\ and\ \bibinfo {author}
  {\bibfnamefont {H.}~\bibnamefont {Shinohara}},\ }\href
  {https://doi.org/10.1063/1.481172} {\bibfield  {journal} {\bibinfo  {journal}
  {The Journal of Chemical Physics}\ }\textbf {\bibinfo {volume} {112}},\
  \bibinfo {pages} {6000} (\bibinfo {year} {2000})},\ \Eprint
  {https://arxiv.org/abs/https://pubs.aip.org/aip/jcp/article-pdf/112/13/6000/19295065/6000\_1\_online.pdf}
  {https://pubs.aip.org/aip/jcp/article-pdf/112/13/6000/19295065/6000\_1\_online.pdf}
  \BibitemShut {NoStop}%
\bibitem [{\citenamefont {Hwang}\ \emph {et~al.}(2012)\citenamefont {Hwang},
  \citenamefont {Siegel}, \citenamefont {Mo}, \citenamefont {Regan},
  \citenamefont {Ismach}, \citenamefont {Zhang}, \citenamefont {Zettl},\ and\
  \citenamefont {Lanzara}}]{hwang2012fermi}%
  \BibitemOpen
  \bibfield  {author} {\bibinfo {author} {\bibfnamefont {C.}~\bibnamefont
  {Hwang}}, \bibinfo {author} {\bibfnamefont {D.~A.}\ \bibnamefont {Siegel}},
  \bibinfo {author} {\bibfnamefont {S.-K.}\ \bibnamefont {Mo}}, \bibinfo
  {author} {\bibfnamefont {W.}~\bibnamefont {Regan}}, \bibinfo {author}
  {\bibfnamefont {A.}~\bibnamefont {Ismach}}, \bibinfo {author} {\bibfnamefont
  {Y.}~\bibnamefont {Zhang}}, \bibinfo {author} {\bibfnamefont
  {A.}~\bibnamefont {Zettl}},\ and\ \bibinfo {author} {\bibfnamefont
  {A.}~\bibnamefont {Lanzara}},\ }\href {https://doi.org/10.1038/srep00590}
  {\bibfield  {journal} {\bibinfo  {journal} {Scientific reports}\ }\textbf
  {\bibinfo {volume} {2}},\ \bibinfo {pages} {590} (\bibinfo {year}
  {2012})}\BibitemShut {NoStop}%
\bibitem [{\citenamefont {Taioli}\ \emph {et~al.}(2016)\citenamefont {Taioli},
  \citenamefont {Gabbrielli}, \citenamefont {Simonucci}, \citenamefont
  {Pugno},\ and\ \citenamefont {Iorio}}]{Taioli2016}%
  \BibitemOpen
  \bibfield  {author} {\bibinfo {author} {\bibfnamefont {S.}~\bibnamefont
  {Taioli}}, \bibinfo {author} {\bibfnamefont {R.}~\bibnamefont {Gabbrielli}},
  \bibinfo {author} {\bibfnamefont {S.}~\bibnamefont {Simonucci}}, \bibinfo
  {author} {\bibfnamefont {N.~M.}\ \bibnamefont {Pugno}},\ and\ \bibinfo
  {author} {\bibfnamefont {A.}~\bibnamefont {Iorio}},\ }\href@noop {}
  {\bibfield  {journal} {\bibinfo  {journal} {Journal of Physics: Condensed
  Matter}\ }\textbf {\bibinfo {volume} {28}},\ \bibinfo {pages} {13LT01}
  (\bibinfo {year} {2016})}\BibitemShut {NoStop}%
\bibitem [{Note1()}]{Note1}%
  \BibitemOpen
  \bibinfo {note} {Extensive analysis of these forces is beyond the scope of
  this paper and will be the subject of subsequent work to be reported
  elsewhere}\BibitemShut {NoStop}%
\bibitem [{\citenamefont {Vozmediano}\ \emph {et~al.}(2010)\citenamefont
  {Vozmediano}, \citenamefont {Katsnelson},\ and\ \citenamefont
  {Guinea}}]{vozmediano2010gauge}%
  \BibitemOpen
  \bibfield  {author} {\bibinfo {author} {\bibfnamefont {M.~A.}\ \bibnamefont
  {Vozmediano}}, \bibinfo {author} {\bibfnamefont {M.}~\bibnamefont
  {Katsnelson}},\ and\ \bibinfo {author} {\bibfnamefont {F.}~\bibnamefont
  {Guinea}},\ }\href@noop {} {\bibfield  {journal} {\bibinfo  {journal}
  {Physics Reports}\ }\textbf {\bibinfo {volume} {496}},\ \bibinfo {pages}
  {109} (\bibinfo {year} {2010})}\BibitemShut {NoStop}%
\bibitem [{\citenamefont {Cortijo}\ and\ \citenamefont
  {Vozmediano}(2007)}]{cortijo2007electronic}%
  \BibitemOpen
  \bibfield  {author} {\bibinfo {author} {\bibfnamefont {A.}~\bibnamefont
  {Cortijo}}\ and\ \bibinfo {author} {\bibfnamefont {M.~A.}\ \bibnamefont
  {Vozmediano}},\ }\href@noop {} {\bibfield  {journal} {\bibinfo  {journal}
  {Europhysics Letters}\ }\textbf {\bibinfo {volume} {77}},\ \bibinfo {pages}
  {47002} (\bibinfo {year} {2007})}\BibitemShut {NoStop}%
\bibitem [{\citenamefont {Muller}\ \emph {et~al.}(1999)\citenamefont {Muller},
  \citenamefont {Schubert},\ and\ \citenamefont {van~de Ven}}]{Muller1999}%
  \BibitemOpen
  \bibfield  {author} {\bibinfo {author} {\bibfnamefont {U.}~\bibnamefont
  {Muller}}, \bibinfo {author} {\bibfnamefont {C.}~\bibnamefont {Schubert}},\
  and\ \bibinfo {author} {\bibfnamefont {A.~E.~M.}\ \bibnamefont {van~de
  Ven}},\ }\href {https://doi.org/10.1023/A:1026718301634} {\bibfield
  {journal} {\bibinfo  {journal} {General Relativity and Gravitation}\ }\textbf
  {\bibinfo {volume} {31}},\ \bibinfo {pages} {1759} (\bibinfo {year}
  {1999})}\BibitemShut {NoStop}%
\end{thebibliography}%
\bibliographystyle{apsrev4-2}

\end{document}